\documentclass[aps,twocolumn,prb,showpacs,superscriptaddress]{revtex4-1}% Physical Review
\usepackage{graphicx}% Include figure files
\usepackage{dcolumn}% Align table columns on decimal point
\usepackage{bm}% bold math
\usepackage{amsmath}% More mathematical features
\usepackage{amssymb}
\usepackage{epsfig}
\usepackage{verbatim} % to make comments
\usepackage{pstricks,pst-grad,pst-plot}% PsTricks
\definecolor{lightyellow}{cmyk}{0,0,0.3,0}  % added by Sergei
\definecolor{lightblue}{cmyk}{0.1,0,0,0}  % added by Sergei
\definecolor{green4}{cmyk}{0.0,0.0,0.7,0.4}  % added by Sergei

\begin{document}

\title{Synergy in spreading processes: from exploitative to explorative foraging
strategies}

\author{Francisco~J.~P{\'e}rez-Reche}
\affiliation{Department of Chemistry, University of Cambridge, Cambridge, UK}
\affiliation{SIMBIOS Centre, University of Abertay,  Dundee,  UK}
\email{p.perezreche@abertay.ac.uk}

\author{Jonathan J. Ludlam}
\affiliation{Churchill College, University of Cambridge, Cambridge,
  UK}

\author{Sergei~N.~Taraskin}
\affiliation{St. Catharine's College and Department of Chemistry,
University of Cambridge, Cambridge, UK}

\author{Christopher~A.~Gilligan}
\affiliation{Department of Plant Sciences, University of Cambridge,
Cambridge,  UK}

\begin{abstract}
An epidemiological model which incorporates synergistic effects that allow the infectivity and/or susceptibility of hosts to be dependent on the number of infected neighbours is proposed.  Constructive synergy induces an exploitative behaviour which results in a rapid invasion that infects a large number of hosts.  Interfering synergy leads to a slower and sparser explorative foraging strategy  that traverses larger distances by infecting fewer hosts. The model can be mapped to a dynamical bond-percolation with spatial correlations that affect the mechanism of spread but do not influence the critical behaviour of epidemics.
\end{abstract}


\maketitle

\date{\today}

%\section{Introduction}
The identification of criteria to predict whether or not a spreading
agent such as an infectious pathogen, rumour or opinion will invade a
population is of great relevance both in biology and social
sciences~\cite{AndersonMay86:invasion,Murray_02:book,Castellano_RMP2009,Centola_Science2010}.
Numerous models have been proposed to gauge the invasiveness of
spreading agents and assess the effectiveness of control in preventing
invasion of
epidemics~\cite{AndersonMay86:invasion,Murray_02:book,Castellano_RMP2009,Gubbins00:thresholds,Keeling01:fmdsci,Dodds_PRL2004,Shao_Havlin_Stanley_PRL2009,Maksim_NatPhys2010,Ben-Jacob_Nature1994,Ferreira_PRE2002}.
Within the context of infectious diseases, much theoretical work has
been done for well-mixed populations  
but
invasion in stochastic, spatially-structured,
individual-based
models~\cite{Keeling99:localinvasion,Grassberger1983,Sander2002,PerezReche_JRSInterface2010,Kenah2007,Miller_PRE2007}
 closer to real
epidemics~\cite{Bailey00:perc,Otten04:perc2,Davis_Nature2008,Pastor-Satorras_Book2004} has also been considered.
These models do not deal with synergistic effects in transmission of
infection thus assuming independent and identical action
between hosts.  
This would suggest that multiple challenges to a
susceptible host from one, two or more neighbouring infected hosts are
independent and not influenced by the local environment.
However, there is evidence for the existence of such effects in
 systems subject to colonisation by fungal and bacterial
pathogens~\cite{Rayner_Mycologia1991_review,Ben-Jacob_Nature1994},  as
well as in tumour growth~\cite{Liotta_Nature2001,Ferreira_PRE2002}.
Synergistic effects have also been experimentally reported in studies of opinion dynamics~\cite{Castellano_RMP2009}, spread of behaviour~\cite{Centola_Science2010}, and animal invasion~\cite{Murray_02:book,Gordon_Book2009}.
The model presented in Ref.~\onlinecite{Dodds_PRL2004} incorporates
some temporal synergistic effects but it deals with well-mixed
populations and spatial synergistic effects are not considered.
Models for opinion dynamics and animal invasion have considered some
constructive synergistic spatial effects (e.g. population
pressure~\cite{Murray_02:book} and social
impact~\cite{Castellano_RMP2009,Centola_PhysicaA2007}).  
However, these effects are too
simple to capture, for instance, possible changes in the foraging
strategies of spreading agents that can significantly affect
important features of  invasions such as their size and time scales.
Here, we present a model for spread of
infection in spatially-structured populations and show that
synergistic effects in transmission of infection have non-trivial and
significant consequences on epidemics.
%
%
%%%%%%%%%%%%%%%%%%
%%%% Figure 1 %%%%
%%%%%%%%%%%%%%%%%%
%%%%%%%%%%%%%%%%%%
%%%% Figure 1 %%%%
%%%%%%%%%%%%%%%%%%
\begin{figure}
\begin{center}
{\includegraphics[clip=true,width=8.0cm]{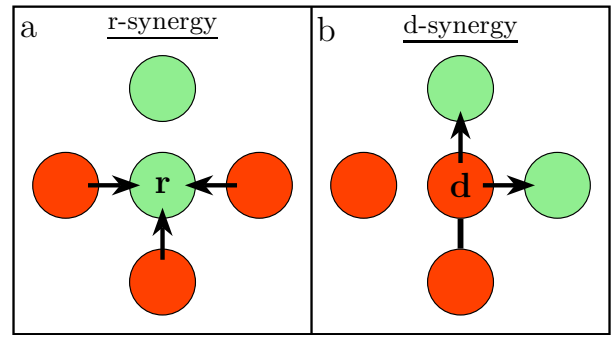}}

\end{center}
\caption{\label{fig:1} Mechanism for (a) r-synergy and (b) d-synergy.
  Dark (red online) and lighter (green online) circles correspond to infected and susceptible hosts,
  respectively.  Arrows indicate possible events for transmission of
  infection.  For r-synergy, the susceptibility of a \emph{recipient}
  susceptible host \textcircled{r} is enhanced (constructive synergy)
  or diminished (interfering synergy) in response to multiple
  simultaneous challenges from two or more neighbouring infected hosts
  (donors). The susceptibility of \textcircled{r} depends on the number
  $n_{\text{r}}$ of neighbours simultaneously challenging it
  ($n_{\text{r}}=3$ in this example).  For d-synergy, the infectivity
  of a \emph{donor} host \textcircled{d} depends upon the number $n_{\text{d}}$ of connections
  of this host to other infectious hosts that can share resources with
  the donor host.  In this example, $n_{\text{d}}=1$ (connection indicated by an edge).  }

\end{figure}

We consider an epidemic spreading through a population of
  susceptible hosts
  placed on the sites/nodes of a square lattice of size $L \times L$.
The infection transmission rate between any donor-recipient (d-r) pair
of hosts depends
on the number of infected hosts
  in the neighbourhood of the d-r pair.
 We focus on
  two particular cases of the model denoted as \emph{r-synergy} and
  \emph{d-synergy} (Fig.~\ref{fig:1}).

%\section{Model}
The model is an extension of a basic spatial model for the SIR epidemic
process in which hosts (sites) can be in one of the three
states~\cite{Grassberger1983}:
susceptible (S), infected (I) or removed and fully immune to
further infection (R).
Once a host is infected, it stays in such a
state for a fixed unit of time, $\tau=1$ and can pass infection during this
infectious period to other S-neighbours, and then it is
removed/recovered (I $\to$ R transition). The infection process (S
$\to$ I transition) occurs through the pathogen being transmitted randomly with rate
$\lambda_{\text{d-r}}(t)$ from one of the I-neighbours (donor) of the
S-host (recipient). Synergistic effects make the rate
$\lambda_{\text{d-r}}(t)$ dependent
%, $\lambda_{\text{d-r}}(t)$,
%to become dependent
on the state of hosts in the neighbourhood of the particular
d-r pair. By definition, $\lambda_{\text{d-r}}(t)$,
is zero before the time of infection of the donor.  Once the donor is
infected, $\lambda_{\text{d-r}}(t)$ becomes positive but, in contrast
to the basic SIR
model~\cite{Grassberger1983,Henkel_Hinrichsen_Book2009}, it can vary during the whole infectious period $\tau$
due to possible changes in the neighbourhood of the 
d-r pair. 
The transmission rate can be conveniently split into two contributions,
\begin{equation}
\lambda_{\text{d-r}}(t)=\max \{0,\alpha+\beta_{\text{d-r}}(t)\}~,
\label{eq:lambda}
\end{equation}
where   $\alpha \geq 0$  is the elementary rate of infection
for  an isolated d-r pair exhibiting no
synergistic effects.
It does not vary  over the infectious period and is
  assumed to be the same for all d-r pairs.
The rate $\beta_{\text{d-r}}(t)$ quantifies the degree of synergy
present and is $\beta_{\text{d-r}}(t)=0$ in the absence of
synergistic effects when the model reduces to the simple SIR process.
The expression for $\beta_{\text{d-r}}(t)$ depends on the type of synergy.
For r-synergy (Fig.~\ref{fig:1}(a)), 
we assume that $\beta_{\text{d-r}}(t)=\beta
(n_{\text{r}}(t)-1)$, where $n_{\text{r}}(t)$ is the number of
neighbours challenging a recipient host.
The rate $\beta$ gives an
effective measure of the strength of synergy which is constructive for
  $\beta>0$ and interfering  if $\beta<0$.
Such a form of $\beta_{\text{d-r}}(t)$ ensures that
$\lambda_{\text{d-r}}=\alpha$ for non-synergistic transmissions with
$n_{\text{r}}=1$.
For d-synergy (Fig.~\ref{fig:1}(b)),  we  assume
  that  $\beta_{\text{d-r}}(t)=\beta n_{\text{d}}(t)$,
where
$n_{\text{d}}(t)$ is the number of infected neighbours connected to a
donor at time $t$ so that $\lambda_{\text{d-r}}=\alpha$ for an isolated infectious
host with $n_{\text{d}}=0$.

\begin{figure}%[h]
\begin{center}

{\includegraphics[clip=true,width=8.0cm]{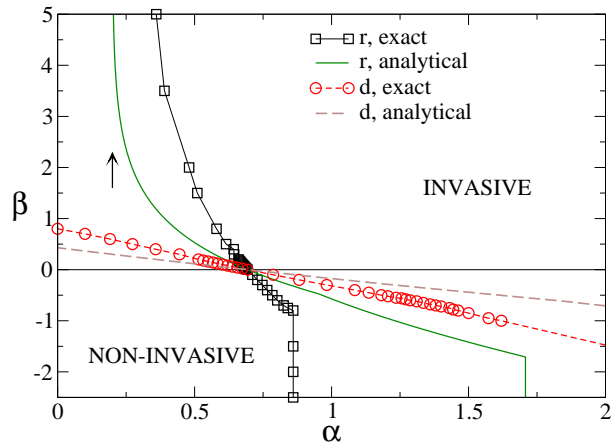}}
\end{center}
\caption{ \label{fig:2_new}
Phase diagram for synergistic epidemics. The
exact    phase boundaries are marked   by squares for
  r-synergy and by circles for d-synergy.
  Epidemics are invasive if the parameters ($\alpha, \beta$)
    are on the right of the phase boundary.
  The arrow indicates the limiting value of $\alpha_{\text{c}}
  \simeq 0.2$ reached asymptotically by the
  phase boundary for r-synergy in the
  limit of large $\beta$ (not
  shown on the scale of the graph).
The approximate phase boundaries obtained analytically by neglecting correlations in transmission of infection 
(cf. Appendix~\ref{sec:phase_diagram})
 are shown for both r- (solid line) and d-synergy (dashed line).
}
\end{figure}

Fig.~\ref{fig:2_new} shows the phase diagram for epidemics starting
from a single infected host placed in the centre of the lattice. The
spread of infection has been numerically simulated by a
\emph{continuous-time} algorithm which is an extension of the
\emph{n-fold} way algorithm~\cite{Fallert_PRE2008}.
The threshold for invasion defines a line of critical points
$\alpha_{\text{c}}(\beta)$ separating the non-invasive regime
where the probability of invasion is
$P_{\text{inv}}=0$ (in an infinite system) from the invasive regime
characterised by $P_{\text{inv}}>0$ \footnote{An epidemic is considered to be invasive if it reaches all
  four edges of the system.}.
%\footnote{An epidemic  is considered
%to be invasive if it  reaches all four edges of the  system.
%The probability of invasion, $P_{\text{inv}}$, is defined as
%the relative number of invasive events out of many (in our
%  simulations, $\gtrsim 5000$)
%stochastic realizations of epidemics (see Sec.~I in \cite{Supplementary}).}.
The phase boundaries
 shown in Fig.~\ref{fig:2_new} correspond to the
values of $\alpha_{\text{c}}(\beta)$ for $L \to \infty$.
The finite-size effects were accounted for and eliminated by
means of finite-size scaling (see Appendices \ref{Sec:Pinv} and \ref{Sec:Crit_FS}).
%

%%%%%%%%%%%%%%%%%
%%% Figure 2   %%
%%%%%%%%%%%%%%%%%
%%%%%%%%%%%%%%%%%
%%% Figure 2   %%
%%%%%%%%%%%%%%%%%

For both types of synergy, $\alpha_{\text{c}}$ is a non-increasing
function of $\beta$, as expected from the monotonic dependence of
$\lambda_{\text{d-r}}$ on $\beta$ [Eq.~\eqref{eq:lambda}].
In the absence of synergy,  the invasion thresholds for
  both types of synergy coincide with
$\alpha_{\text{c}}(\beta=0)= \tau^{-1}\ln 2$ (cf. Fig.~\ref{fig:2_new} and Appendix~\ref{Sec:Pinv}).
The larger deviations of $\alpha_{\text{c}}(\beta)$ from $\alpha_{\text{c}}(0)$ 
for a given $\beta$
observed for d-synergy are  due to the fact that, except for the
initially infected
  host, d-synergy is present in every
transmission event with $\beta \neq 0$ because $n_{\text{d}} \geq 1$
for at least some time.
In contrast, for r-synergy to
be operative in a transmission event, there must be $n_{\text{r}} \geq
2$ attacking neighbours which is not
necessarily the case in every transmission event.

For r-synergy with large positive values of $\beta$, the critical line $\alpha_{\text{c}}(\beta)$ tends towards the
limiting value
$\alpha_{\text{c}}(\infty) \simeq 0.2$ (Fig.~\ref{fig:2_new}).
In this situation,
$\lambda_{\text{d-r}}=\alpha < \infty$ for $n_{\text{r}}=1$ but
$\lambda_{\text{d-r}}=\infty$ if $n_{\text{r}}>1$, meaning that hosts
being simultaneously attacked by more than one neighbour are infected
immediately.
In cases with large interference in transmission (i.e.
very  negative
$\beta$), the  invasion threshold  is located at $\alpha_{\text{c}}= 0.86\pm 0.01$
independently of the value of $\beta$. In this regime, $\lambda_{\text{d-r}}$ does not depend on
$\beta$  and corresponds to the limiting situation with
$\lambda_{\text{d-r}}=\alpha$ for $n_{\text{r}}=1$ and
$\lambda_{\text{d-r}}=0$ for any $n_{\text{r}}>1$.
\begin{comment}
In this regime,
invasion occurs only through single-neighbour infective challenge
  with elementary rate
$\alpha$ since
interfering transmission for sites being challenged by more than one
neighbour is forbidden.
\end{comment}

%\subsubsection{d-synergy}
For d-synergy and values of $\beta \gtrsim 0.8$, invasion is possible
for any positive $\alpha$.  The condition $\alpha>0$ is necessary in
order for the epidemic to start from a single inoculated site.  Once
the pathogen is transmitted to one of the neighbours of the initially
inoculated host, $n_{\text{d}}=1$ for the newly infected host. The
combination of $n_{\text{d}}=1$ and synergy is sufficient to make
invasion possible, irrespective of the value of $\alpha$.

\begin{figure}%[h]
\begin{center}
{\includegraphics[clip=true,width=8cm]{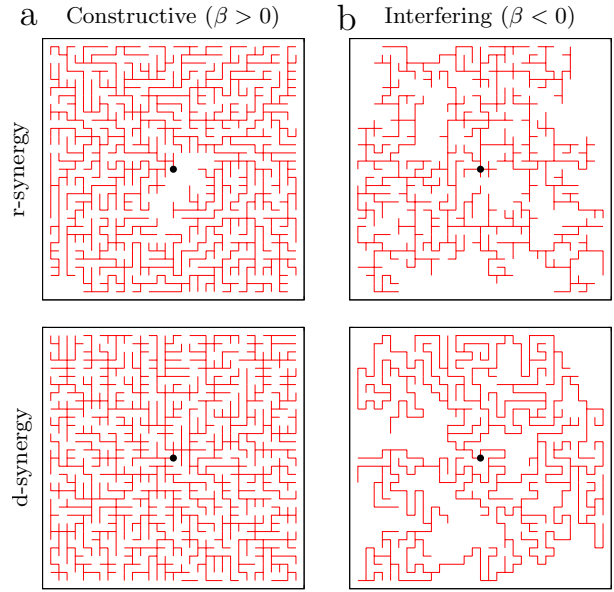}}

\end{center}
\caption{ \label{fig:patterns} Patterns of infection. Illustration of the effect of (a)
  constructive and (b) interfering synergy on the
  patterns of infection
  spreading from the central host (marked by solid circle) in systems of
  linear size $L=31$. The synergy rates are
  $\beta=5$ and $\beta=-5$ for
  patterns with constructive and interfering synergy,
  respectively.
  All snapshots correspond to the final state of the
  epidemic with only R and S hosts remaining.
  Solid lines indicate those edges
  between d-r pairs that have transmitted the pathogen at some time
  during the course of the epidemic.
  The value of $\alpha$ has been
  chosen in each case such
  that the probability of invasion is
  $P_{\text{inv}}=0.5$ for all snapshots: (a) $\alpha= 0.47$ for r-synergy and $\alpha= 0.18$ for d-synergy. (b) $\alpha= 0.81$ for r-synergy and $\alpha= 5.22$ for d-synergy.}
\end{figure}

Systems with constructive synergy are characterised by dense patterns
of invasion (Fig.~\ref{fig:patterns}(a)).  In contrast, the invaded
region is more sparse for interfering synergy
(Fig.~\ref{fig:patterns}(b)).  This scenario applies to both d- and
r-synergy.
Despite the fact that patterns for interfering synergy are
more sparse, it is striking that epidemics with interfering synergy
can nevertheless be as invasive as epidemics with constructive synergy
in terms of the spatial extent of infection.
For instance, all the
patterns shown in Fig.~\ref{fig:patterns} have the same probability of
invasion, $P_{\text{inv}}=0.5$.
In terms of the mean spatial density  of
invasion defined as the relative number of hosts infected by the
pathogen before invasion occurs with a given probability, we conclude  that the
larger the interference in transmission of the pathogen, the less damaging
the invasion is (Appendix \ref{Sec:Space_Eff}).
This result can
be qualitatively understood as follows.  
Interfering synergy favours transmission of infection to
hosts with few infected neighbours and disfavours transmission to
hosts with several previously infected neighbours. Therefore,
infection has a tendency to evolve towards poorly infected regions
rather than infecting as many hosts as possible. 
These mechanisms are qualitatively similar for both types of synergy but the density of infection is larger for d-synergy than for r-synergy for any $\beta \neq 0$ (see Appendix \ref{Sec:Space_Eff} for more detail).

\begin{figure}%[h]
\begin{center}
\begin{pspicture}(-0.2,0)(8,3.0)
%\rput(2.5,2.5){\Large{(a)}}
%\psset{unit=1.0cm}
\vspace{0.2cm}
\rput(1.8,1.3){{\includegraphics[clip=true,height=3.5cm]{d-synergy_Phase_Diagram_Invasion_Branching.eps}}}
\rput(0.0,2.9){\large{a}}

%-------------------------------------
%\rput(5.4,3.0){\underline{d-synergy}}
\rput(6.5,1.3){{\includegraphics[clip=true,height=3.0cm]{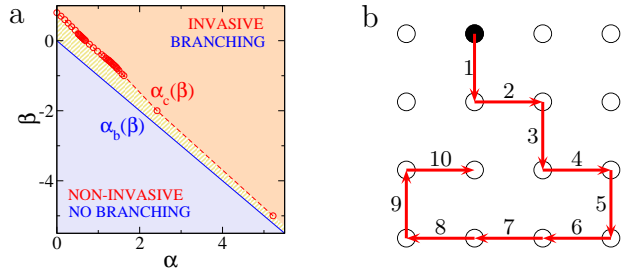}}}
\rput(4.7,2.9){\large{b}}

%\psgrid
\end{pspicture}

\end{center}
\caption{ \label{fig:mechanisms} Branching and invasion transitions for d-synergy. (a)
  Diagram showing the branching transition line
  $\alpha_{\text{b}}(\beta)=-\beta$ together with the
phase boundary for invasion/non-invasion transition,
$\alpha_{\text{c}}(\beta)$.  Branching is forbidden in
  the region under the continuous line where $\alpha<\alpha_{\text{c}}(\beta)$. Invasive epidemics
  can occur in the region above the dashed line corresponding to $\alpha>\alpha_{\text{c}}(\beta)$ where branching is present.
The intermediate region between the continuous and dashed lines
  with $\alpha_{\text{b}}<\alpha<\alpha_{\text{c}}$ corresponds to epidemics that
  display branching but are not invasive.
(b) Example of path for
  infection with d-synergy in the regime without branching ($\alpha \leq
  \alpha_{\text{b}}$). Infection starts
  from the host marked by a
  solid circle and evolves along the path indicated by
  arrows. Arrow numbers define
  schematically
  a sequence of infection events. 
 For
the first infected host (solid circle), $n_{\text{d}}=0$ and infection
can be transmitted to any of its neighbours at a rate
$\lambda_{\text{d-r}}=\alpha$.  However, $n_{\text{d}}=1$ as soon as
the pathogen is transmitted to one of the neighbours (arrow 1) and
thus $\lambda_{\text{d-r}}=0$ because $\alpha \leq -\beta$.  At this
moment, transmission of infection is arrested until the initially
infected host recovers.  After this recovery, $n_{\text{d}}=0$ for the
newly infected host and infection can be transmitted to one of its
nearest neighbours at rate $\alpha$.  Iteration of this process over
time gives growing SAWs. The state after the
event marked by arrow 10 illustrates the phenomenon of self-trapping.
}
\end{figure}

For constructive synergy, the patterns of invasion are not very much
influenced by the particular type of synergy.
For 
interfering synergy, the degree of branching of the paths followed by
the pathogen is clearly smaller for d-synergy than for r-synergy
(cf. panels in Fig.~\ref{fig:patterns}(b)).
Branching is always possible for r-synergy.
In  contrast, the patterns of invasion for d-synergy display a branching
transition for a value of $\alpha=\alpha_{\text{b}}(\beta)=-\beta$:
branching occurs for $\alpha>\alpha_{\text{b}}(\beta)$ but it is
absent for $\alpha \leq \alpha_{\text{b}}(\beta)$
(Fig.~\ref{fig:mechanisms}(a)). In the
later case, the trajectories of invasion are of the type
followed by a growing self-avoiding walk
(SAW)~\cite{Pietronero_PRL1985,Hemmer_PRA1986}
with an example shown in Fig.~\ref{fig:mechanisms}(b).
  As expected for growing SAWs, the pathogen can display
self-trapping, meaning that the epidemic stops if the infection
reaches a host surrounded by hosts that have already been infected
(Fig.~\ref{fig:mechanisms}(b)).
For values of $\alpha > \alpha_{\text{b}}(\beta)$
(Fig.~\ref{fig:patterns}(b), lower panel), 
$\lambda_{\text{d-r}}=\alpha+\beta$ is positive for
$n_{\text{d}}=1$ and branching is possible.

%%%%%%%%%%%%%%%%%
%%% Figure 3   %%
%%%%%%%%%%%%%%%%%
%%%%%%%%%%%%%%%%%
%%% Figure 3   %%
%%%%%%%%%%%%%%%%%
%

%%%%%%%%%%%%%%%%%
%%% Figure 4   %%
%%%%%%%%%%%%%%%%%
%%%%%%%%%%%%%%%%%
%%% Figure 4   %%
%%%%%%%%%%%%%%%%%
%
%

As
a measure of the \emph{temporal efficiency} for invasion,
we consider the time, $t_{\text{inv}}$, it takes for the pathogen to
invade the system.  Numerical simulations (see Appendix~\ref{Sec:t_eff})
show that for any given value of $P_{\text{inv}}$, the time
$t_{\text{inv}}$ decreases with increasing $\beta$ (for both types of
synergy). Therefore, systems with interfering synergy are less
time-efficient than those with constructive synergy. 
The largest
deviations of $t_{\text{inv}}$ from its value without
synergy are obtained for epidemics with d-synergy that operates in every transmission event.

%\section{Discussion}

The analysis presented above demonstrates that synergy in
  transmission of infection has significant and sometimes paradoxical
  and unexpected effects on epidemics.
Despite the simple
  assumptions of the model (such
  as, e.g., short-range synergy and linear dependence
  of the rate of
  infection on the number of infected nearest neighbours), it reproduces
  explorative and
  exploitative foraging strategies that are typically observed in
bacterial, fungal, and tumour growth~\cite{Rayner_Mycologia1991_review,Ferreira_PRE2002,Ben-Jacob_Nature1994}.
  The
    explorative/exploitative behaviour in our model is linked to interfering/constructive
    synergy. The foraging strategy adopted by fungi, bacteria or ants is known to be
    explorative/exploitative when resources are limited/abundant.

Changes in the foraging strategy are ultimately due to the
spatial correlations in transmission rates emerging as a consequence of
synergy.
Spatial correlations in $\lambda_{\text{\text{d-r}}}$ appear
because the
neighbourhood of sufficiently close pairs of hosts have common nodes
and thus the rates for each pair are not
mutually independent.
Synergistic epidemics can then be regarded as a
  \emph{correlated} dynamical percolation analogous to the well-known mapping of
  the non-synergistic SIR process to uncorrelated dynamical
percolation (see ~\cite{Grassberger1983,Henkel_Hinrichsen_Book2009,Sander2002,PerezReche_JRSInterface2010} and 
  details on the mapping to correlated dynamical percolation in Appendix~\ref{Sec:Corr_Dyn_Perc}).
In most situations, spatial correlations in transmission are
short-ranged
and the critical behaviour of epidemics at $\alpha_{\text{c}}$ belongs to the dynamical uncorrelated
bond-percolation
universality class~\cite{Grassberger1983,Henkel_Hinrichsen_Book2009}. However,
for large interfering d-synergies with $\alpha<\alpha_{\text{b}}(\beta)$
(region under continuous line in Fig.~\ref{fig:mechanisms}(a)) correlations become
effectively long-ranged and epidemics are growing SAWs
whose critical properties belong to the universality class of
the standard SAW~\cite{Pietronero_PRL1985,Hemmer_PRA1986}. 
 Although the SAW behaviour affects the local properties of epidemics with $\alpha \gtrsim
\alpha_{\text{b}}(\beta)$, the large-scale behaviour at $\alpha_{\text{c}}$ is not affected (i.e. the critical exponents  at invasion are the same as those for uncorrelated bond percolation, as shown in detail in Appendix~\ref{Sec:Crit_FS}). 
This is a consequence of the fact that the probability that a
growing SAW invades a large system is known to be
zero due to the self-trapping
phenomenon~\cite{Pietronero_PRL1985,Hemmer_PRA1986}. This implies that
$\alpha_{\text{c}}>\alpha_{\text{b}}$ for any value of $\beta$ (cf. Fig.~\ref{fig:mechanisms}(a)).

The proposed model becomes analytically tractable if spatial correlations in transmission are assumed to be negligible. The analytic solutions provide a good qualitative description of the main features of the phase diagram for both r- and d-synergy. Correlations in the exact model prevent the approximate description from being quantitative (cf. Appendix~\ref{sec:phase_diagram} for a complete description).

In summary, the presented work shows that
synergistic effects at the individual level play an important role in
invasion at the population level. The analysis has been restricted to the spread of epidemics in 2D regular networks relevant for, e.g., populations of plants in a field.  The extension of our work to synergistic effects for spreading processes in higher-dimensional lattices or more complex networks is not only conceptually appealing but also important for  problems in multiple disciplines~\cite{Note4}. Although such an extension is technically straightforward, understanding the interplay between synergistic effects and topological properties for different types of networks is a challenging task for future work.
\begin{acknowledgments}
  We thank G.J. Gibson and W. Otten for helpful discussions and funding from BBSRC (Grant No. BB/E017312/1). CAG acknowledges support of a BBSRC Professorial Fellowship.
\end{acknowledgments}

\section*{Appendices}
\appendix

\section{ Probability of invasion}
\label{Sec:Pinv}
\begin{figure*}
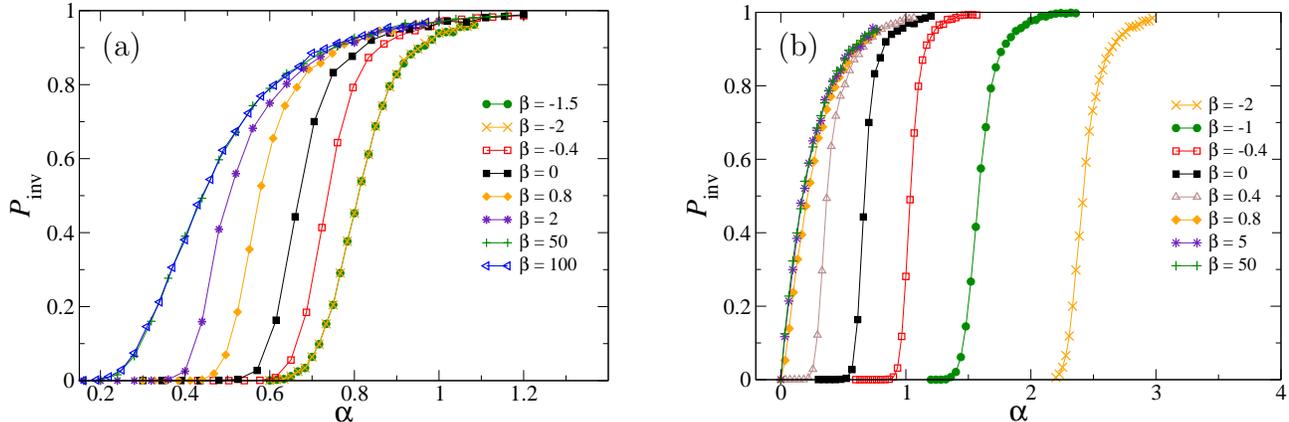
%[h]
\begin{center}

\begin{pspicture}(0,3)(17,8)

\rput(4,5){{\includegraphics[clip=true,width=8cm]{Pinv_sy1_alpha_L31_PBC.eps}}}
\rput(1.5,7.2){\large{(a)}}
\rput(13,5){{\includegraphics[clip=true,width=8cm]{Pinv_sy3_alpha_L31_PBC.eps}}}
\rput(10.5,7.2){\large{(b)}}

%\psgrid

\end{pspicture}
%%%%%%%%%%%%%%%%%%%%%%%%%%%%%%%%%%%%%%%%%%%%%%%%%%%%%%%%%%%%%%%%%%%%%%%%

\end{center}
\caption{ \label{fig:Pinv_alpha}
%Old writing
Probability of invasion $P_{\text{inv}}$ in a system of
  linear size $L=31$ as a function
  of the elementary rate $\alpha$ for (a) r-synergy and (b)
  d-synergy. Different curves correspond to different values of the
  synergy rate, $\beta$, as indicated in the legend for each case.
}
\end{figure*}

In this section, we discuss the dependence of the probability of
invasion, $P_{\text{inv}}$, on the rates $\alpha$ and $\beta$. 
The probability of invasion is defined as the relative number of
  invasive events out of many (in our simulations, $\gtrsim 5000$)
  stochastic realizations of epidemics.   
Although we will deal with
systems of finite size, the trends of $P_{\text{inv}}$ (see
Fig.~\ref{fig:Pinv_alpha}) are in qualitative agreement with those
expected from the phase diagram for epidemics in infinite systems
shown in the main text (Fig.~2).  For both r- and d-synergy, the
invasion curves ($P_{\text{inv}}$ \emph{vs} $\alpha$) approach zero
for small transition rates $\alpha$ thus describing the non-invasive
regime for epidemics.  For large values of $\alpha$, the invasion
probability is finite so that it describes the invasive regime.  The
invasion curves start to deviate from zero at progressively lower
values of $\alpha$ as $\beta$ increases.  This illustrates the
non-increasing character of $\alpha_{\text{c}}(\beta)$ for increasing
$\beta$ discussed in the main text. The tendencies of $P_{\text{inv}}$
in extreme situations require further explanations that are different
for r- and d-synergy.

%%%%%%%%%%%%%%%%%%%%%%%%%%%%%%%%%%%%%%%%%%%%%%%%%%%%%%%%%%%%
%% Figure: Pinv vs \alpha for different values of \beta   %%            
%%%%%%%%%%%%%%%%%%%%%%%%%%%%%%%%%%%%%%%%%%%%%%%%%%%%%%%%%%%%

\paragraph{\bf r-synergy}
Fig.~\ref{fig:Pinv_alpha}(a) shows the probability of invasion for
epidemics with r-synergy in systems of linear size $L=31$
as a function of $\alpha$ (invasion curves) for several values of $\beta$.
For large positive values of $\beta$, the invasion curves
tend to a single (master) curve which can be described by
a limiting function of $\alpha$ that does not depend significantly on $\beta$.
This is illustrated in Fig.~\ref{fig:Pinv_alpha}(a) by two curves
for $\beta=50$ and $\beta=100$ coinciding within numerical error.
Consequently, on the phase diagram,
the critical line $\alpha_{\text{c}}(\beta)$ tends towards the
limiting value
$\alpha_{\text{c}}(\infty) \simeq 0.2$ (shown
by  the arrow in Fig.~2, main text).
In cases with large interference in transmission (corresponding to
very  negative
$\beta$), the curves for $P_{\text{inv}}(\alpha)$
again collapse on a limiting curve
shown in Fig.~\ref{fig:Pinv_alpha}(a) for $\beta=-1.5$ and
$\beta=-2.0$.
The corresponding invasion threshold in the
thermodynamic limit is located at $\alpha_{\text{c}}= 0.86\pm 0.01$
independently of the value of $\beta$.

\paragraph{\bf d-synergy}
For d-synergy and values of $\beta \gtrsim 0.8$, invasion is possible for any
positive value of $\alpha$.
In this regime, the invasion curves collapse on a
master curve  that does not depend on the
value of $\beta$
(cf. curves in Fig.~\ref{fig:Pinv_alpha}(b) for $\beta=5$ and
$\beta=50$).
For negative values of $\beta$, the invasion curves do not
approach  a
limiting function as $\beta$ decreases
(Fig.~\ref{fig:Pinv_alpha}(b)).
As a result, $\alpha_{\text{c}}$ increases
monotonically with decreasing $\beta$ (see Fig.~2, main text).

\section{Critical behaviour and finite-size effects}
\label{Sec:Crit_FS}
In this section, we give details about the
methods used for obtaining the phase diagram for invasion in the limit
$L \rightarrow \infty$ (see Fig.~2 in the main text). 
In addition, we give
numerical support for the statements made in the main text about the
critical behaviour displayed by epidemics with d-synergy in the
different regions presented in Fig.~4(a) of the main text.

The data points for the phase
boundary, $\alpha_{\text{c}}(\beta)$, in 
the phase diagram for invasion were
 obtained by analysing both the
probability of invasion, $P_{\text{inv}}$, and the average relative
number, $N_1$, of epidemics spanning the system in one and only one
direction (1D-spanning epidemics). 
The analysis of $N_1$ presented
here is similar to that performed for avalanches in spin models with
quenched disorder \cite{PerezReche2003,PerezReche_PRL2008}.  

Below we
show that, for epidemics with constructive or weakly interfering
synergy, the method based on the analysis of 1D-spanning clusters
gives more accurate estimates for
$\alpha_{\text{c}}(\beta)$ than those obtained by using
$P_{\text{inv}}$. 
In particular, our estimates for the invasion threshold for non-synergistic epidemics based on $N_1$ are in excellent agreement with the value $\alpha _{\protect \text {c}}(0)= \tau^{-1}\ln 2$ expected from the mapping to bond-percolation
  valid for $\beta =0$  \footnote{For hosts on a square lattice, the mapping of the SIR model to bond-percolation gives the critical
  transmissibility $T_{\text {c}}= 1-e^{-\tau \alpha_{\text {c}}(0) }=1/2$ \cite{Stauffer1994} (cf. Eq.~\eqref{eq_SI:T_maintext}) and thus $\alpha _{\protect \text {c}}(0)=\tau^{-1}\ln 2$.}.
In contrast, for epidemics with large interfering
d-synergy, the estimation of $\alpha_{\text{c}}(\beta)$ based on
$P_{\text{inv}}$ is more accurate than that achieved by using $N_1$.

\subsection{Epidemics with constructive
or weakly interfering synergy}
\begin{figure*}
\begin{center}
\begin{pspicture}(-0.2,-0.3)(17,5)

\rput(1.5,4.3){\large{a}}
\rput(4,2.0){{{\includegraphics[clip=true,width=8.5cm]{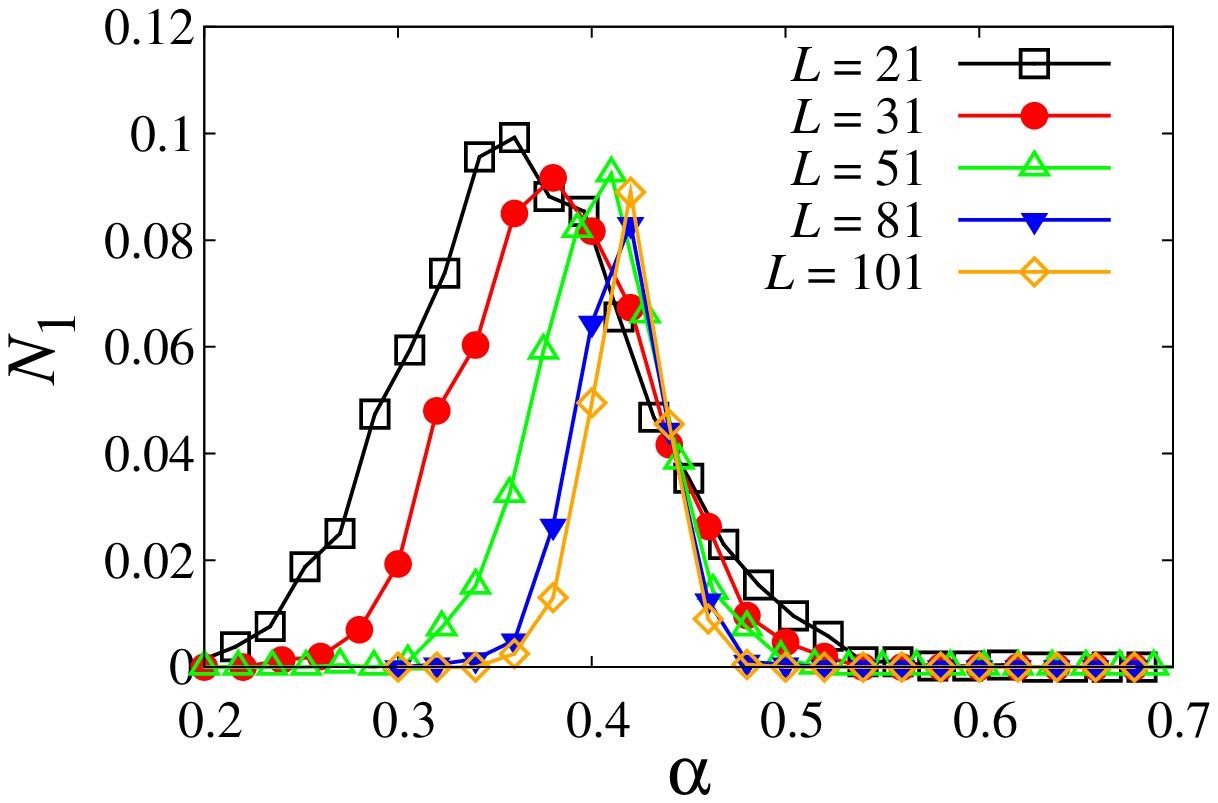}}}}
\rput(10.5,4.3){\large{b}}
\rput(13,2.0){{{\includegraphics[clip=true,width=8.5cm]{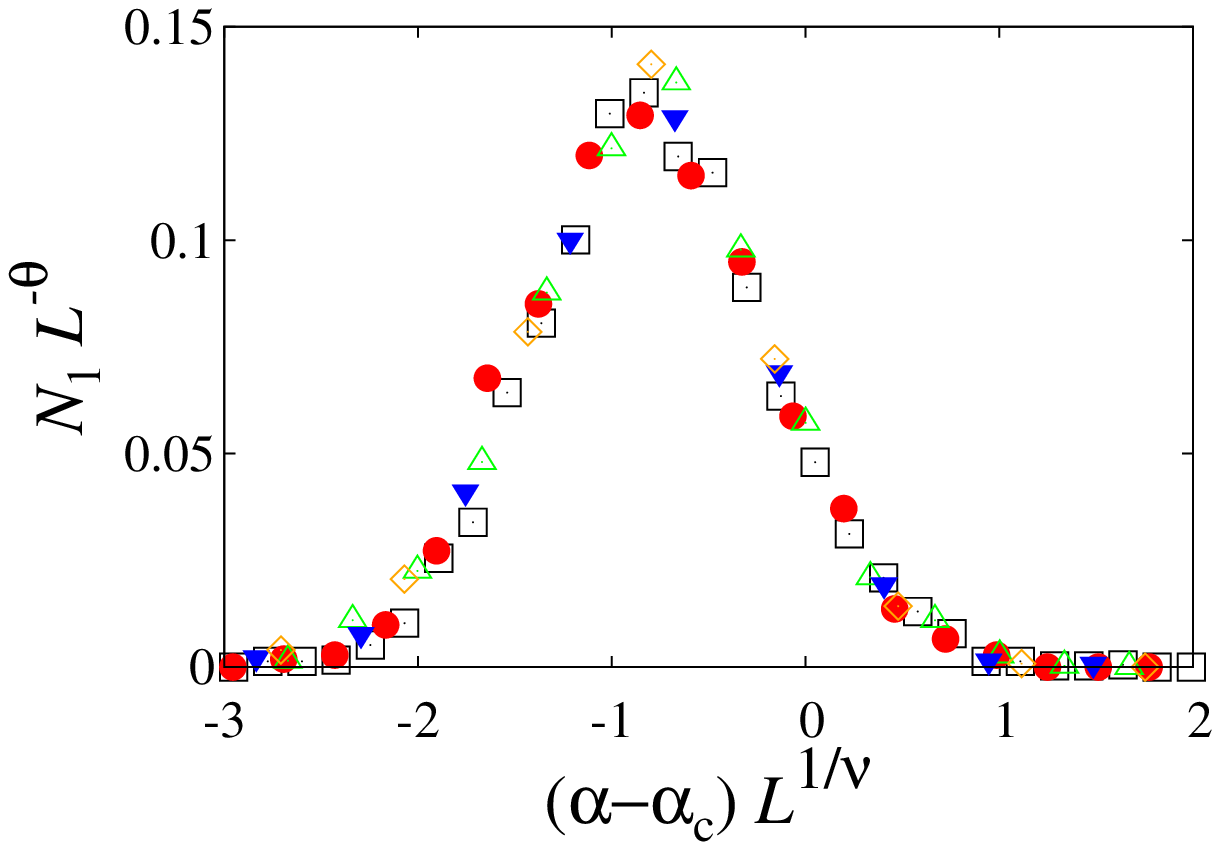}}}}

%\psgrid
\end{pspicture}

\end{center}
\caption{\label{fig:SI_N1_be_0c3} (a) Dependence of the average
  relative  number of 
  1D-spanning epidemics, $N_1$, on the elementary rate $\alpha$ for
  d-synergy with $\beta=0.3$. Different curves correspond to systems
  of different size $L$, as marked in the legend. (b) Scaling collapse of $N_1$ according to the
  scaling law \eqref{eq_SI:N1_Scaling} for the curves shown in
  (a). The critical exponents have been set to the values
  $\nu=4/3$ and $\beta=5/36$, corresponding to values found for
  uncorrelated 
  percolation. 
  The scaling collapse gives the
  invasion threshold $\alpha_{\text{c}} = 0.445 \pm 0.005$.}
\end{figure*}

For epidemics with constructive or weakly interfering synergy, the
quantity $N_1$ has a peak when plotted as a function of $\alpha$ for
a fixed value of $\beta$. 
This is the case for both r- and
d-synergy. 
Fig.~\ref{fig:SI_N1_be_0c3} shows a representative example corresponding to
constructive d-synergy with $\beta=0.3$.
The position of the peak for $N_1$ is related
 to the critical value of the parameter $\alpha$ separating
   invasive and non-invasive regimes.
  Indeed, in the non-invasive
  regime, $N_1 \simeq 0$ because the clusters of removed hosts are
   of small size and thus they have a
   negligible probability to touch the boundaries
of a finite system.
  On the
  other hand, $N_1$ is expected to be close to zero in the
  invasive regime as 
  well because the pathogen spreads in two directions rather than
  along one only.
  As a result, $N_1$ can be finite at the
  invasion threshold
  or, due to finite-size effects, in a certain region around the
  threshold. 
  The critical value $\alpha_{\text{c}}(\beta)$ in the limit $L
  \rightarrow \infty$ is obtained using  the scaling
  properties of $N_1(\alpha,L)$ in the vicinity of the  invasion threshold.
  Due to the absence of characteristic length scales at
  criticality, the dependence of $N_1$ on $\alpha$ and $L$ is expected
  to obey the following scaling law \cite{PerezReche2003,PerezReche_PRL2008}:
\begin{equation}
\label{eq_SI:N1_Scaling}
N_1(\alpha,L)=L^{\theta} \tilde{N}_1((\alpha-\alpha_{\text{c}})L^{1/\nu})~,
%\tag{S.1}
\end{equation}
where $\theta$ and $\nu$ are critical exponents and $\tilde{N}_1$ is
a
scaling function which depends on $\alpha$ and $L$ through the product
$(\alpha-\alpha_{\text{c}})L^{1/\nu}$ only.  
The values of $\theta$, $\nu$, and
$\alpha_{\text{c}}$ can be determined by scaling
collapse for $N_1(\alpha,L)$. 
Technically, this can be achieved by plotting the quantity 
$L^{-\theta} N_1$ \emph{vs} $(\alpha-\alpha_{\text{c}})L^{1/\nu}$ and requiring
that the scaling hypothesis \eqref{eq_SI:N1_Scaling} is satisfied
(i.e. curves for different $L$ collapse on a single master curve
corresponding to $\tilde{N}_1$).  
Fig.~\ref{fig:SI_N1_be_0c3}(b) shows an
example of such scaling collapse for d-synergy with $\beta=0.3$. 
In this case, the collapse gives the values $\alpha_{\text{c}} = 0.445 \pm 0.005$
and $\theta=-0.10 \pm 0.05$. 
Bearing in mind the mapping of SIR
epidemics to dynamical uncorrelated bond-percolation holding for
$\beta=0$~\cite{Grassberger1983}, the value of $\nu$ has been fixed to
$\nu=4/3$ corresponding to uncorrelated percolation
\cite{Stauffer1994}. The good quality of the collapse suggests that,
despite the existence of correlations in transmission of infection for
$\beta \neq 0$, the critical behaviour at
$\alpha_{\text{c}}(\beta)$ falls into the 
universality class for the uncorrelated percolation. 
This is the expected
behaviour when correlations are short-ranged \cite{Isichenko1992}. 
\begin{figure*}%[h]
\begin{center}
\begin{pspicture}(-0.2,-0.3)(17,5)

\rput(1.5,4.3){\large{a}}
\rput(4,2.0){{{\includegraphics[clip=true,width=8.5cm]{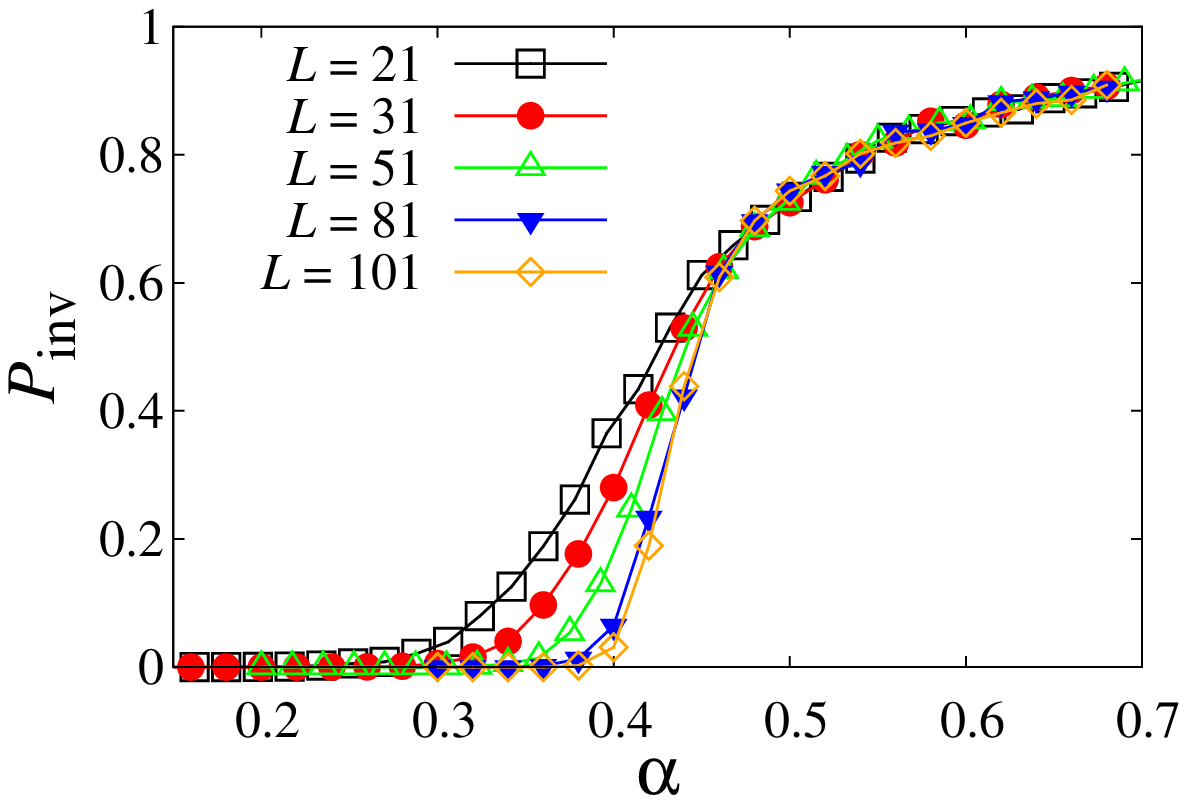}}}}
\rput(10.5,4.3){\large{b}}
\rput(13,2.0){{{\includegraphics[clip=true,width=8.5cm]{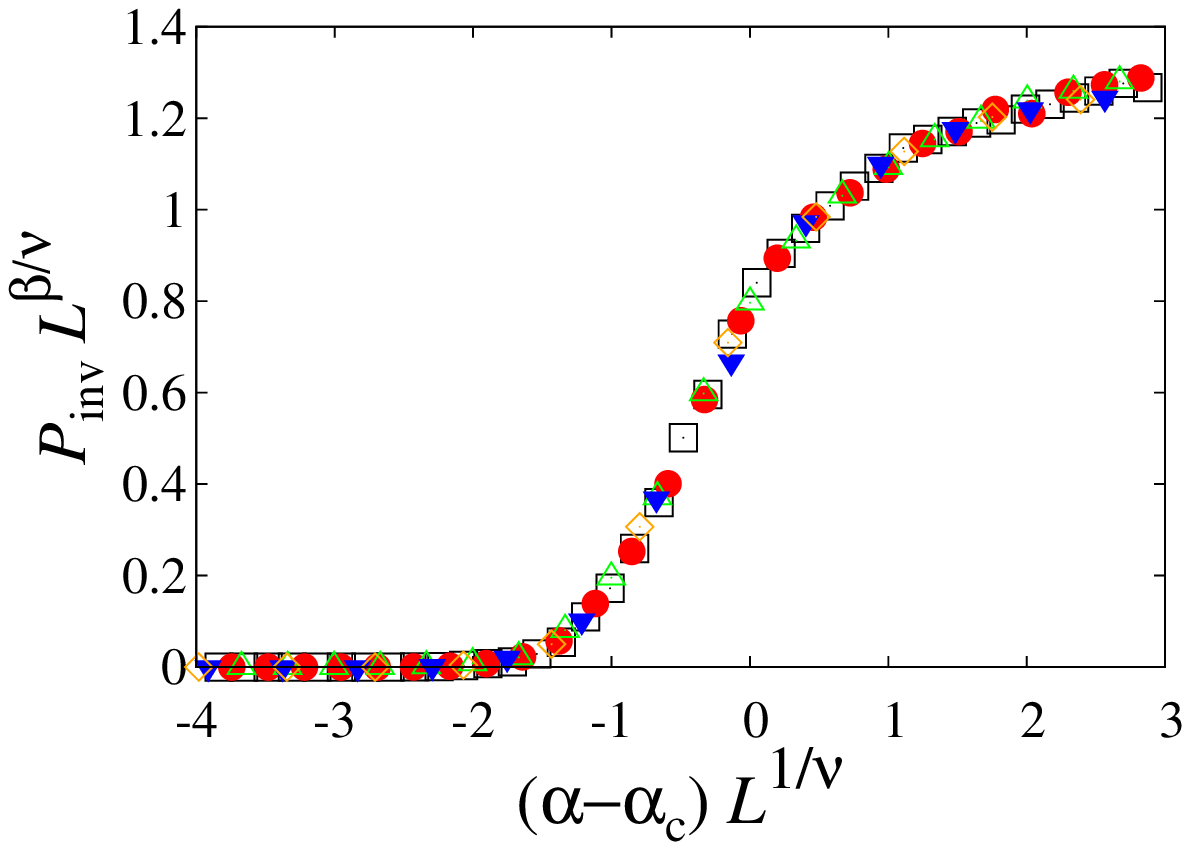}}}}

%\psgrid
\end{pspicture}

\end{center}
\caption{\label{fig:SI_Pinv_be_0c3} (a) Dependence of the probability
  of invasion, $P_{\text{inv}}$, on the elementary rate $\alpha$ for 
  d-synergy with $\beta=0.3$. Different curves correspond to systems
  of different size $L$, as marked in the legend. 
  (b) Scaling collapse of $P_{\text{inv}}$ according to the
  scaling law given by Eq.~\eqref{eq_SI:Pinv_Scaling} for the
  curves shown in 
  (a). The critical exponents have been set to the values
  $\nu=4/3$ and $\beta=5/36$. The invasion threshold has been set to
  the value $\alpha_{\text{c}} = 0.445$ obtained from the collapse of
  $N_1$ (cf. Fig.~\ref{fig:SI_N1_be_0c3}).}
\end{figure*}

The above
statement is also supported by the behaviour of 
$P_{\text{inv}}$ in the vicinity of the invasion threshold which
satisfies the following scaling hypothesis,
\begin{equation}
\label{eq_SI:Pinv_Scaling}
P_{\text{inv}}=L^{-\beta/\nu}
\tilde{P}_{\text{inv}}((\alpha-\alpha_{\text{c}})L^{1/\nu}),
%\tag{S.2}
\end{equation}
with the exponents $\nu=4/3$ and $\beta=5/36$ corresponding to
uncorrelated percolation. 
As an example,
Fig.~\ref{fig:SI_Pinv_be_0c3}(b) shows the scaling collapse for the
curves shown in
Fig.~\ref{fig:SI_Pinv_be_0c3}(a) for several system sizes. 
The collapse in Fig.~\ref{fig:SI_Pinv_be_0c3}(b) has been obtained by
setting $\alpha_{\text{c}}$ to the value $\alpha_{\text{c}} = 0.445$
obtained from the collapse of $N_1$. 
The quality of this collapse is remarkably good despite the fact
that no free parameters have been used (i.e. the value of $\nu$,
$\beta$, and $\alpha_{\text{c}}$ has been considered as being fixed). 
%\newpage

If instead of fixing $\alpha_{\text{c}}$, we try to estimate its value
from the collapse of 
$P_{\text{inv}}$, the sigmoidal shape of $P_{\text{inv}}$ prevents
the
estimate from being as accurate as the one obtained from the collapse
of $N_1(\alpha,L)$ which are peak-shaped.

\subsection{Epidemics with large interfering d-synergy}

\begin{figure*}%[h]
\begin{center}
\begin{pspicture}(-0.2,-0.3)(17,5)

\rput(1.5,4.3){\large{a}}
\rput(4,2.0){{{\includegraphics[clip=true,width=8.5cm]{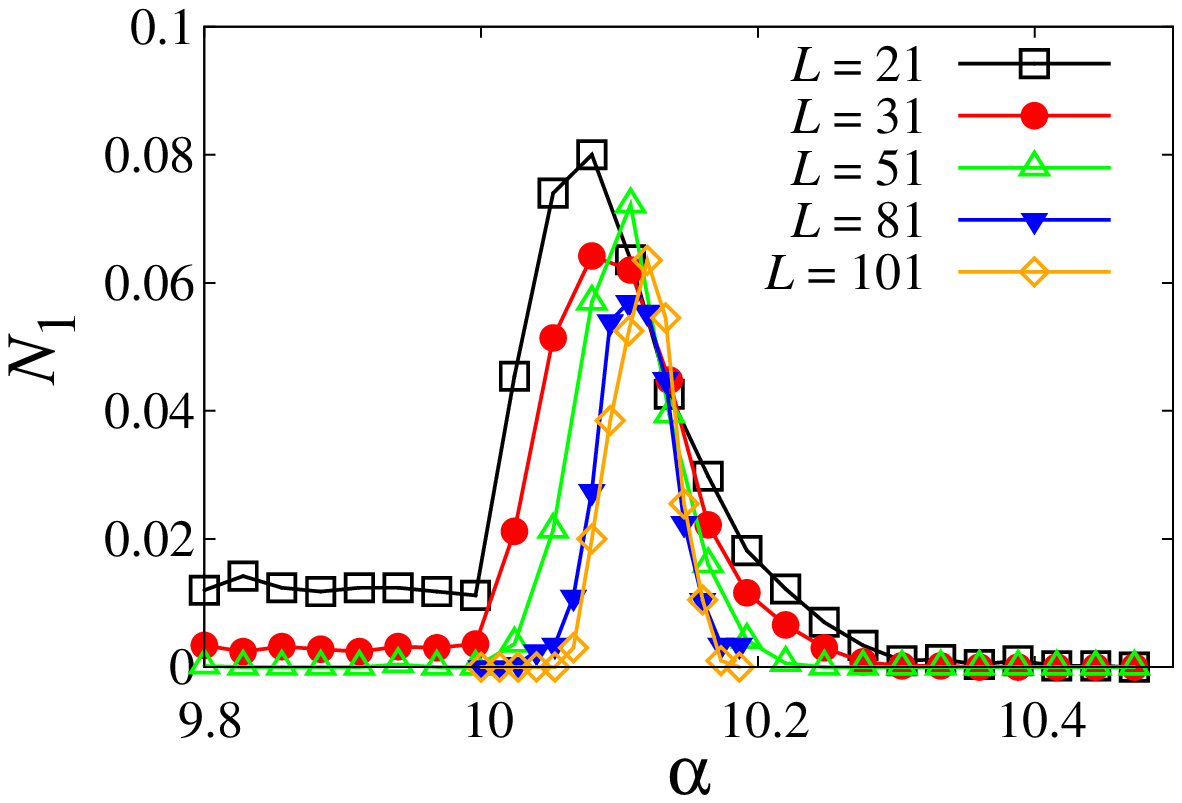}}}}
\rput(10.5,4.3){\large{b}}
\rput(13,2.0){{{\includegraphics[clip=true,width=8.5cm]{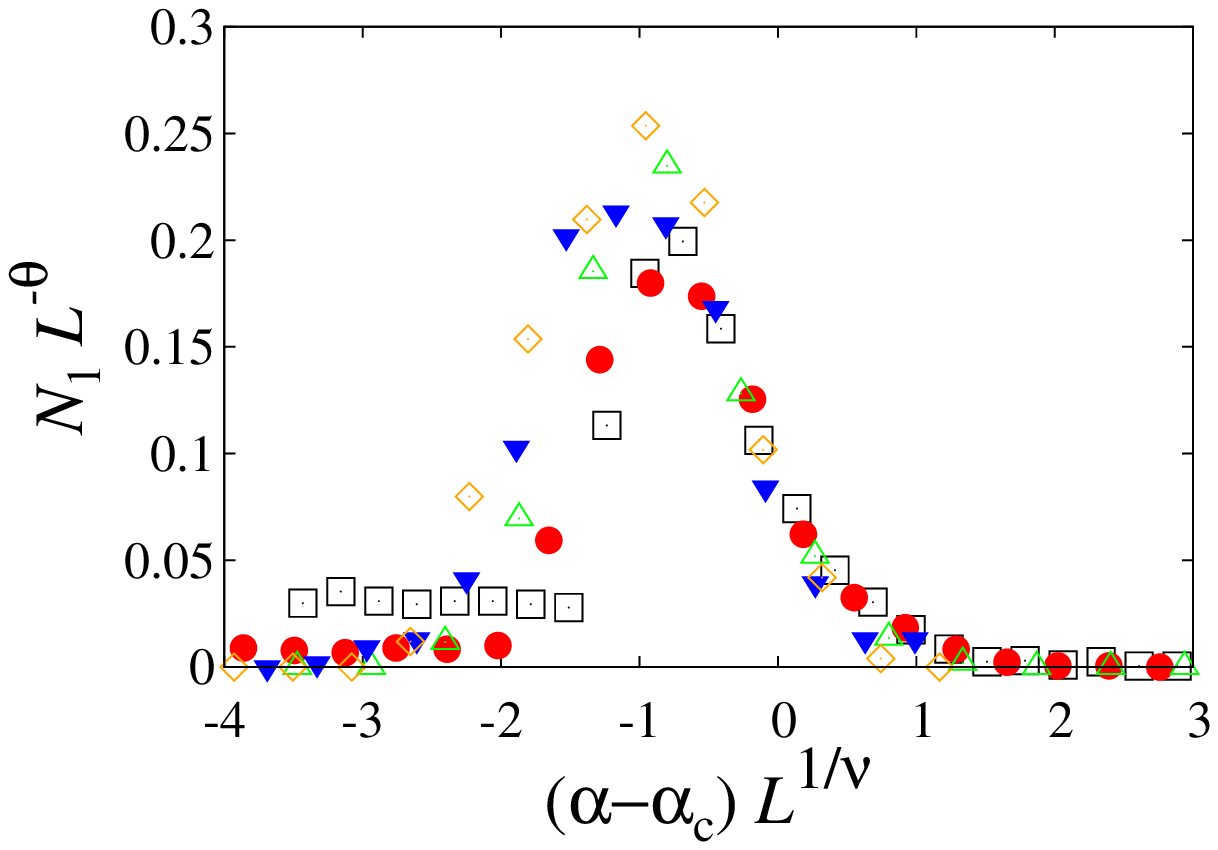}}}}

%\psgrid
\end{pspicture}

\end{center}
\caption{\label{fig:SI_N1_be_m10c0} (a) Dependence of the average
  relative  number of 
  1D-spanning epidemics, $N_1$, on the elementary rate $\alpha$ for
  d-synergy with $\beta=-10.0$. Different curves correspond to systems
  of different size $L$, as marked in the legend. (b) Scaling collapse of $N_1$ according to the
  scaling law given by Eq.~\eqref{eq_SI:N1_Scaling} for the
  curves shown in (a). The critical exponents have been set to the values
  $\nu=4/3$ and $\beta=5/36$. The invasion threshold has been set to
  the value $\alpha_{\text{c}} = 10.15$ obtained from the collapse of
  $P_{\text{inv}}$ (cf. Fig.~\ref{fig:SI_Pinv_be_m10c0}).
}
\end{figure*}

In the main text, we have shown that epidemics with interfering
d-synergy behave as growing self-avoiding walks (SAWs) for $\alpha
\leq \alpha_{\text{b}}=-\beta$ (cf. Fig.~4(a), main
text). 
Fig.~\ref{fig:SI_N1_be_m10c0}(a) shows $N_1$
\emph{vs} $\alpha$ for epidemics with $\beta=-10$ spreading in systems of different size. 
As can be seen,
growing SAWs give a constant contribution to $N_1$ for $\alpha
\leq  10 =\alpha_{\text{b}}(\beta)$. 
The probability that a growing SAW spans
the system in one dimension decreases with the system size, $L$, and
thus the contribution of such objects to $N_1$ tends to zero as
$L \to \infty$. 
In our simulations, the systems are finite and contribution
of the 1D SAWs to $N_1$ is not negligible.
As a consequence, the scaling collapses of
$N_1$ based on the hypothesis  by
  Eq.~\eqref{eq_SI:N1_Scaling} are of low 
quality and not very useful for the estimation of
$\alpha_{\text{c}}(\beta)$ for some negative values of
  $\beta$.

\begin{figure*}
\begin{center}
\begin{pspicture}(-0.2,-0.3)(17,5)

\rput(1.5,4.3){\large{a}}
\rput(4,2.0){{{\includegraphics[clip=true,width=8.5cm]{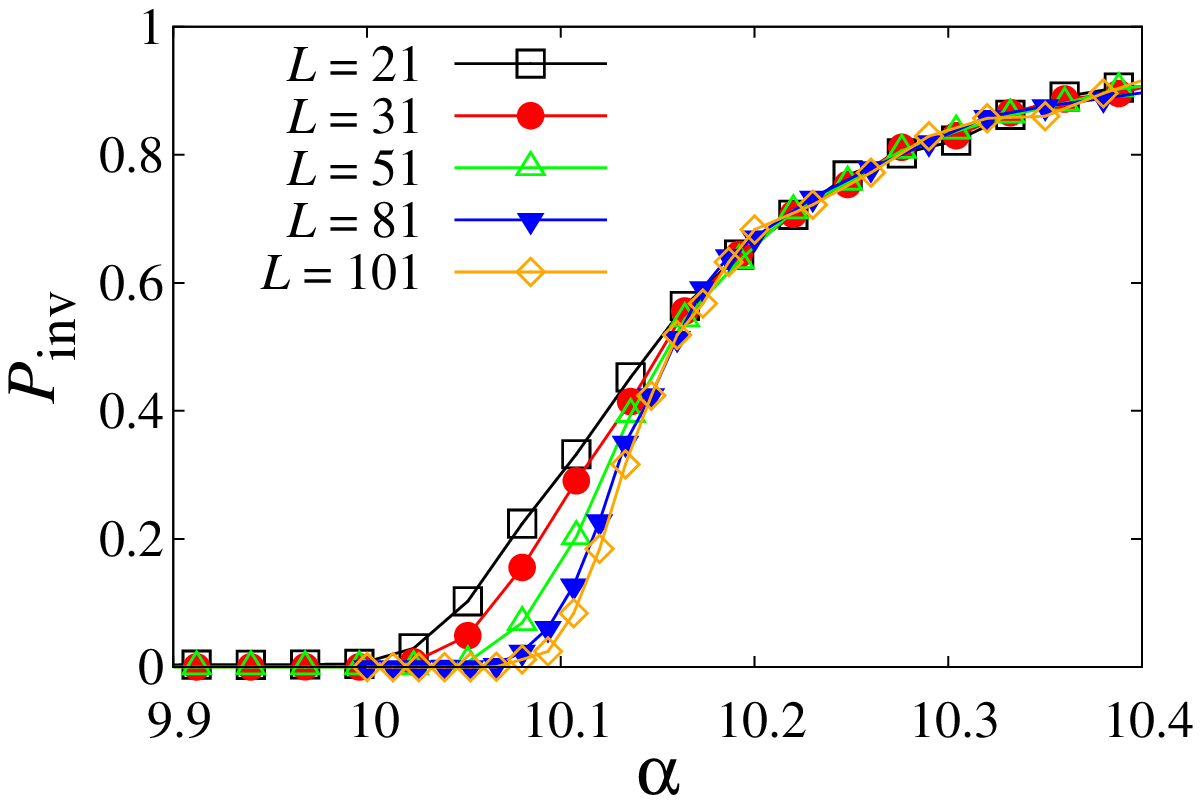}}}}
\rput(10.5,4.3){\large{b}}
\rput(13,2.0){{{\includegraphics[clip=true,width=8.5cm]{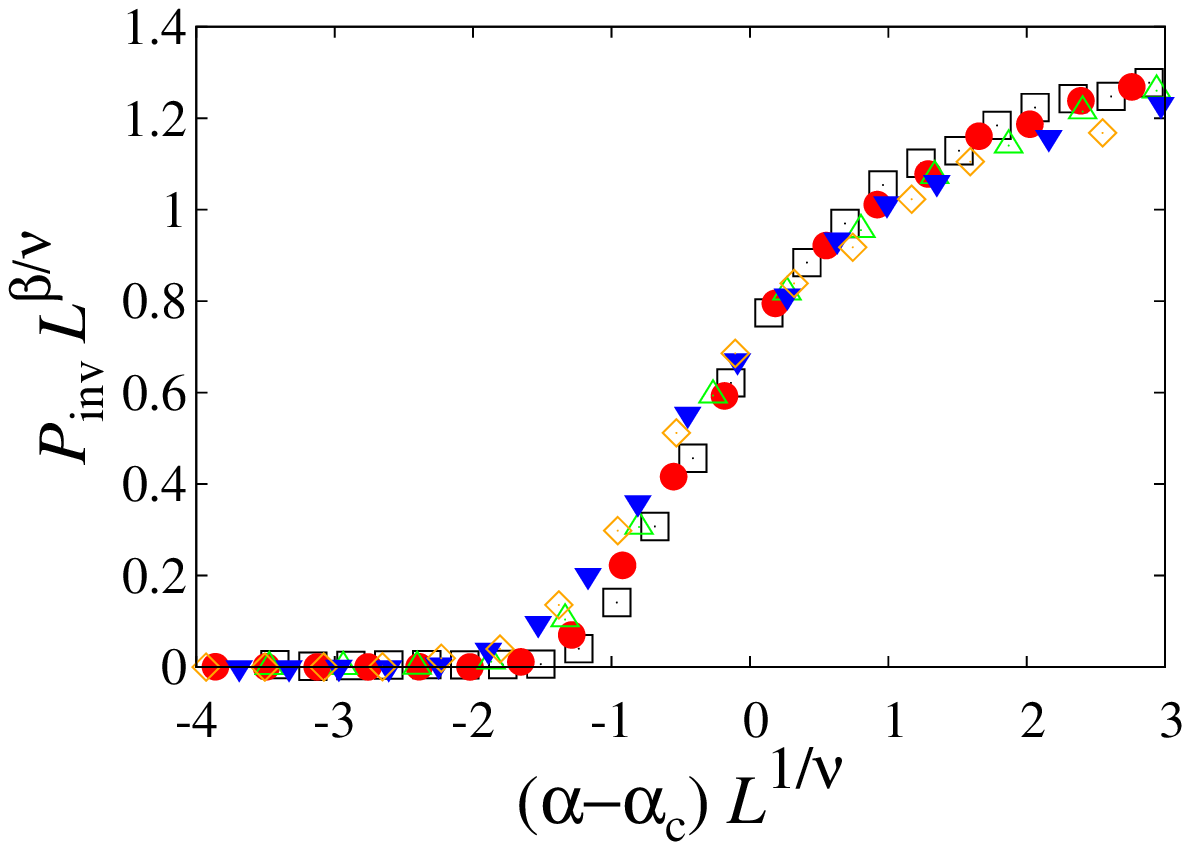}}}}

%\psgrid
\end{pspicture}

\end{center}
\caption{\label{fig:SI_Pinv_be_m10c0} (a) Dependence of the
  probability of invasion, $P_{\text{inv}}$, on the elementary rate
  $\alpha$ for d-synergy with $\beta= -10$. 
  Different curves correspond to systems
  of different size $L$, as marked in the legend. (b) Scaling collapse
  of $P_{\text{inv}}$ according to the 
  scaling law given by Eq.~\eqref{eq_SI:Pinv_Scaling} for the
  curves shown in (a).
  The critical exponents have been set to the values
  $\nu=4/3$ and $\beta=5/36$, corresponding to uncorrelated
  percolation. The scaling collapse gives the
  invasion threshold $\alpha_{\text{c}} = 10.15 \pm 0.03$.}
\end{figure*}

In this situation, the estimates of the critical values of the rate
  $\alpha$ are more conveniently obtained by analysing the probability of
  invasion, $P_{\text{inv}}$. 
This quantity is less affected by the growing SAW epidemics because invasion
requires reaching all the four edges of the
system. 
Fig.~\ref{fig:SI_Pinv_be_m10c0}(a) shows the dependence of
$P_{\text{inv}}$ on $\alpha$ for
$\beta=-10$. 
Fig.~\ref{fig:SI_Pinv_be_m10c0}(b) demonstrates  the scaling collapse of
the curves displayed in panel (a) with the exponents $\nu=4/3$ and $\beta=5/36$
corresponding to uncorrelated percolation and
$\alpha_{\text{c}}=10.15 \pm 0.03$. 
The quality of the collapse suggests that,
despite the proximity of $\alpha_{\text{c}}$ to $\alpha_{\text{b}}$,
the behaviour at large scales corresponds to that of dynamical
uncorrelated percolation. 
Use of the same value of the critical rate,
  $\alpha_{\text{c}}=10.15$, for $N_1(\alpha,L)$ results in a collapse of much poor quality (see Fig.~\ref{fig:SI_N1_be_m10c0}).
 As expected, the quality of the
collapse is reasonable only for relatively large
values of $(\alpha-\alpha_c)L^{1/\nu}$ for which the influence of 
growing SAWs is negligible.

\section{Mean spatial density of invasion}
\label{Sec:Space_Eff}
\begin{figure}%[h]
\begin{center}
{\includegraphics[clip=true,width=8cm]{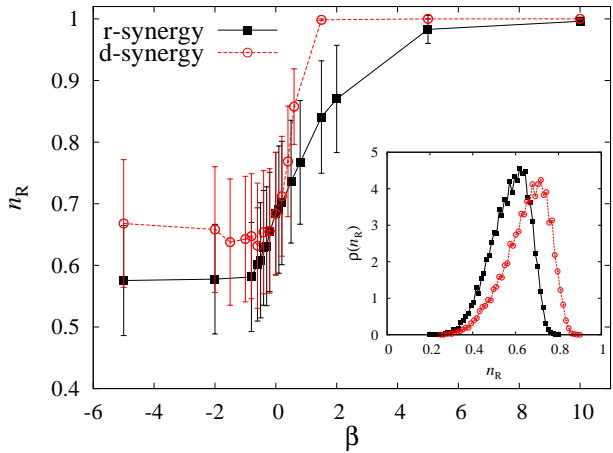}}

\end{center}
\caption{\label{fig:Spatial_Eff} Mean density of invasion, $n_{\text{R}}$, as a
  function of $\beta$ for epidemics with $P_{\text{inv}}=0.5$ in
  systems of linear size $L=31$. Symbols (squares
  for r-synergy and  circles for
  d-synergy) correspond to the mean of $n_{\text{R}}$ over stochastic realisations and error bars
  show the standard deviation.
  The inset displays representative examples of the probability density
  function for the mean density of invasion, $\rho(n_{\text{R}})$, for
  $\beta = -2$.
}
\end{figure}

In this section, we give further quantitative support to the results
discussed in the main text concerning the effect of synergy on the
spatial density of invasion. 
The mean spatial density of invasion is defined as the relative number of 
hosts, $n_{\text{R}}$, that are in the removed state 
(R) by the end of an invasive epidemic (i.e. the relative number of hosts that have been infected during the course of the epidemic and are removed by the end).
Due to stochasticity in the transmission of infection, different
realisations of invasive epidemics characterised by the same
parameters $\alpha$ and $\beta$ have a different \emph{random} value 
for $n_{\text{R}}$.  
The probability density function for the density of infection, $\rho(n_{\text{R}})$, has a single peak for any values of
the parameters $\alpha$ and $\beta$ (see the inset in
Fig.~\ref{fig:Spatial_Eff}).  
Therefore, $n_{\text{R}}$ has a
well-defined scale that, due to the small degree of asymmetry of
$\rho(n_{\text{R}})$, can be properly represented by the mean $\langle
n_{\text{s}} \rangle$ with dispersion given by the standard deviation.
Fig.~\ref{fig:Spatial_Eff} shows the dependence of the spatial
efficiency on $\beta$ for invasive epidemics with value of $\alpha$
chosen in each case so that $P_{\text{inv}} = 0.5$. 
As stated in the
main text, the mean density of invasion exhibits a global tendency to increase with
increasing $\beta$ both for r- and d-synergy.

As expected, $n_{\text{R}}$ is identical for the two types of
synergy if $\beta=0$. For any non-zero value of $\beta$, the density of infection for d-synergy, $n_{\text{R,d}}$, is larger than that for r-synergy, $n_{\text{R,r}}$. In cases with constructive synergy, it is likely that $n_{\text{R,d}} \geq n_{\text{R,r}}$ because, as argued in the main text, synergistic
effects are more prominent for d-synergy (they
operate in every transmission event). 
A plausible explanation for the origin of the inequality $n_{\text{R,d}} > n_{\text{R,r}}$ for $\beta<0$ can be given by recalling that branching is more frequent in paths of infection for r-synergy than for d-synergy. 
Due to the higher degree of branching for r-synergy, it is more probable that hosts become effectively isolated from infection for this type of synergy if they are simultaneously challenged by several neighbours which interfere and do not transmit infection. As an extreme case, consider a situation in which a host is simultaneously challenged by its all four neighbours. It is clear that the challenged host will become inaccessible forever if infection is not transmitted by any of the four challenging neighbours. 
In contrast, the lower degree of branching for d-synergy makes the existence of effectively isolated hosts less likely. As a consequence, more hosts can be infected for d-synergy during the course of epidemics.

\section{Temporal efficiency}
\label{Sec:t_eff}
\begin{figure}%[h]
\begin{center}
\includegraphics[clip=true,width=8cm]{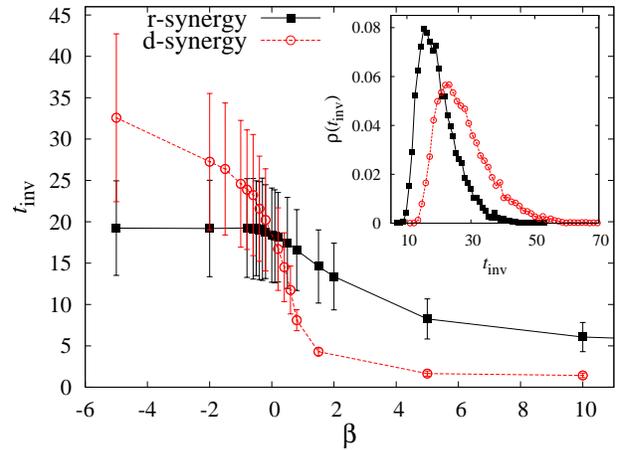}

\end{center}
\caption{ \label{fig:tinv} Time to invasion, $t_{\text{inv}}$, as a
  function of $\beta$ for epidemics with $P_{\text{inv}}=0.5$ in
  systems of linear size $L=31$. Symbols (squares
  for r-synergy and  circles for
  d-synergy) correspond to the mean of $t_{\text{inv}}$ and error bars
  show the standard deviation.
  The inset displays representative examples of the probability density
  function for the time to invasion, $\rho(t_{\text{inv}})$, for
  $\beta = -2$.}
\end{figure}

This section complements the part of the main text devoted to the time
to invasion, $t_{\text{inv}}$. 
Due to stochasticity in the
transmission of infection, $t_{\text{inv}}$ is a random variable
described by a probability density function, $\rho(t_{\text{inv}})$,
which has a single peak for any
  values of the parameters $\alpha$ and $\beta$ (see the inset in
  Fig.~\ref{fig:tinv}). 
We can then proceed in an analogous manner as
  we have done above for the 
mean density of infection and describe  
  $t_{\text{inv}}$ by its mean $\langle
  t_{\text{inv}} \rangle$ and dispersion given by the standard deviation.
Fig.~\ref{fig:tinv} shows that for both types of
synergy $\langle t_{\text{inv}} \rangle$ decreases with increasing
$\beta$.
The dispersion of $\rho(t_{\text{inv}})$ also decreases with increasing $\beta$, as indicated
by the error bars in Fig.~\ref{fig:tinv}.

By definition, the time to invasion is identical for the two types of synergy if $\beta=0$. 
For $\beta \neq 0$, the largest
deviations of $t_{\text{inv}}$ from its value without
synergy ($\beta=0$) (cf. circles and squares in
  Fig.~\ref{fig:tinv}) are for epidemics exhibiting d-synergy. This is mostly due to the fact that d-synergy operates in every transmission event. In average, this makes transmission events slower/quicker for interfering/constructive d-synergy. In addition to this factor, it is likely that the higher degree of branching in the foraging strategy for infection with interfering r-synergy also contributes to making  $t_{\text{inv}}$ smaller for r-synergy with $\beta<0$.

%%%%%%%%%%%%%%%%%%%%%%%%%%%%%%%%%%%%
%%% Figure. Time to invasion      %%
%%%%%%%%%%%%%%%%%%%%%%%%%%%%%%%%%%%%
%%%%%%%%%%%%%%%%%%%%%%%%%%%%%%%%%%%%%

\section{Invasion as a correlated dynamical
             bond-percolation problem}
\label{Sec:Corr_Dyn_Perc}
The aim of this section is twofold.
First, we give detailed
definitions for the transmissibility in synergistic epidemics and
related quantities such as its probability density function (p.d.f.) and
mean value.
Second, the effect of correlations on the probability of
invasion summarised in the main text is analysed here in more detail.

\subsection{Transmissibility}

In the presence of synergistic effects, the transmission of infection
can be described as a non-homogeneous Poisson process with the
time-dependent infection rate $\lambda_{\text{d-r}}(t)$ defined
by Eq.~(1) in  the main text.
The probability that the infection has not
been transmitted in a d-r (donor-recipient) pair by time $t$
defines the survival probability $S_{\text{d-r}}(t)$.
For a non-homogeneous Poisson process, $S_{\text{d-r}}(t)$
satisfies the following differential equation
\cite{Ludwig_MBiosc1975,Pellis_MBiosc2008}:
\[
\frac{\text{d}S_{\text{d-r}}}{\text{d}t}=-\lambda_{\text{d-r}}(t)
S_{\text{d-r}}(t)~.
\]
The solution of this equation with initial condition $S(t=0)=1$
(such a condition ensures that the pathogen
is not  transmitted instantaneously when the donor is infected)
is
\begin{equation}
\label{eq_SI:Survival}
S_{\text{d-r}}(t)=\exp \left[-\int\limits_0^t
  \lambda_{\text{d-r}}(t)~\text{d}t \right]~.
%\tag{S.3}
\end{equation}

The transmissibility $T_{\text{d-r}}$ is defined as the probability
that the pathogen is transmitted
from the donor to the recipient
over the infectious period of the
donor, $\tau$.
Therefore, it can be expressed in terms of the
survival probability  as follows:
\begin{equation}
\label{eq_SI:T}
T_{\text{d-r}}=1-S_{\text{d-r}}(\tau)~.
%\tag{S.4}
\end{equation}
Substitution of the expression for $S_{\text{d-r}}$ given by
Eq.~\eqref{eq_SI:Survival}
into Eq.~\eqref{eq_SI:T} results in 
the following expression for $T_{\text{d-r}}$:
\begin{equation}
\label{eq_SI:T_maintext}
T_{\text{d-r}}=1-\exp \left( - \int_0^{\tau} \lambda_{\text{\text{d-r}}}(t)\, \text{d}t
\right)~.
%\tag{S.5}
\end{equation}

In the absence of synergistic effects, the rate $\lambda_{\text{d-r}}(t)=\alpha$ remains constant over the infectious period $\tau$ and is the same for all d-r pairs. In this case, the transmissibility reduces to the homogeneous value, $T_{\text{d-r}}=1-e^{-\tau \alpha}$, that plays a central role in the mapping of SIR epidemics to the well-known uncorrelated dynamical percolation \cite{Grassberger1983,Henkel_Hinrichsen_Book2009}. In this mapping, $T_{\text{d-r}}$ is identified with the bond probability and $P_{\text{inv}}$ from an initially inoculated site is identified with the probability $P_{\infty}$ that such site belongs to the infinite cluster of connected sites in the dynamical percolation problem\cite{Stauffer1994}. For SIR epidemics, $P_{\text{inv}}$ is fully parameterised by $T_{\text{d-r}}$. This is analogous to the fact that $P_{\infty}$ is fully parameterised by the bond probability in dynamical percolation.

When synergy is present, the transmissibility $T_{\text{d-r}}$  is a functional that depends on the time evolution of the rate $\lambda_{\text{d-r}}(t)$ which, in turn, depends on the infection history of the neighbouring hosts to the d-r pair. 
Each d-r pair involved in an epidemic is in general characterised by a different dependence of the rate $\lambda_{\text{d-r}}(t)$ on time. 
As a consequence, the field of transmissibilities is spatially heterogeneous. In addition, the dependence of $T_{\text{d-r}}$ on the neighbourhood of the d-r pair introduces non-trivial correlations in transmission and thus in transmissibilities.

In order to study the role of synergy-induced correlated heterogeneity at the host level
on $P_{\text{inv}}$ we proceed in a way inspired from previous works
dealing with heterogeneous SIR
epidemics~\cite{kuulasmaa1982,Cox1988,Kenah2007,Miller_JApplProbab2008,Neri_JRSInterface2010,Handford_JRSInterface2010}.  
The simplest situation with heterogeneity in transmission corresponds to epidemics where $T_{\text{d-r}}$ are independent random variables for all d-r pairs. In this case, $P_{\text{inv}}$ only depends on the mean transmissibility, $\langle T \rangle$ \cite{Sander2002,PerezReche_JRSInterface2010}. 
In more complicated situations,  the transmissibilities for different d-r pairs are not independent and $P_{\text{inv}}$ is a function of the whole set of transmissibilities, $\{T_{\text{d-r}}\}$, that cannot be completely parametrised by $\langle T \rangle$~\cite{kuulasmaa1982,Cox1988,Kenah2007,Miller_JApplProbab2008,Neri_JRSInterface2010,Handford_JRSInterface2010}. 
An exact mapping of such SIR epidemics to uncorrelated percolation is not possible in general. 
However, the use of $\langle T \rangle$ can still be useful to analyse the consequences that heterogeneity in local transmission has on $P_{\text{inv}}$. For instance, a considerable progress has been made in understanding the role of correlations in epidemics where heterogeneity in transmission is associated with heterogeneity in recovery times of infected hosts. 
In this case, the following important result has been rigorously derived~\cite{kuulasmaa1982,Cox1988,Neri_JRSInterface2010}: for a given value of $\langle T \rangle$, the resilience to invasion increases with increasing degree of heterogeneity (more precisely, $P_{\text{inv}}^{\text{het}}(\langle T \rangle) \leq P_{\text{inv}}^{\text{hom}}(\langle T \rangle)$, where $P_{\text{inv}}^{\text{het}}$ and $P_{\text{inv}}^{\text{hom}}$ are the probabilities of invasion for heterogeneously and homogeneously distributed removal times, respectively). 
Here, we show that the dependence of $P_{\text{inv}}$ on $\langle T \rangle$ for synergistic epidemics is more complicated (see Fig.~\ref{fig:PinvTavg_r-syn}).  
In spite of that, in subsections \ref{sub-sec:r-synergy} and \ref{sub-sec:d-synergy} we show that analysing the dependence of $P_{\text{inv}}$ on the degree of synergy for given $\langle T \rangle$ is still informative.

For synergistic epidemics we define the mean transmissibility in terms 
of two averages: spatial average in each realisation of epidemics and
stochastic over different epidemic realisations. 
The spatial average for transmissibility,  $\overline{T}_r$,
is calculated for a particular $r$-th  realisation of the epidemic
 in the following manner,
\[
\overline{T}_r=\frac{1}{N_{\text{d-r}}}\sum_{\text{d-r}}T_{\text{d-r}}~,
\]
where the sum extends over the number $N_{\text{d-r}}$  of d-r pairs
in the final state of the epidemic, i.e. over all d-r pairs
challenged by the infection. 
The value of $T_{\text{d-r}}$ in the above equation for each d-r pair is calculated using Eq.~\eqref{eq_SI:T_maintext} with the transmission rate 
$\lambda_{\text{d-r}}(t)$ measured numerically for the $r$-th realisation of epidemic. 
The direct numerical evaluation of the transmissibility for synergistic epidemics as a frequency of successful transmission of infection between donor and recipient in the d-r pair would require reproduction of time-dependent transmission rates giving exactly the same integral over time, $-\int_{0}^{\tau}\lambda_{\text{d-r}}(t) \text{d}t$ (i.e. all possible rates $\lambda_{\text{d-r}}(t)$ giving the same value of $T_{\text{d-r}}$ in Eq.~\eqref{eq_SI:T_maintext}). This imposes a very restrictive condition on the time of infection of the nodes in the neighbourhood of the d-r pair and thus evaluation of $T_{\text{d-r}}$ as a frequency can be hardly a feasible computational task.
The mean transmissibility, $\langle T \rangle$, is obtained by stochastic averaging of $\overline{T}_r$ for $R$ different realisations of the epidemic:
\[
\langle T \rangle=\frac{1}{R}\sum_{r=1}^R \overline{T}_r~,
\]
where the value of  $\overline{T}_r$ is averaged over $R$
  different stochastic realisations of the epidemic. 
Note that averaging over stochastic realisations is necessary to account for the fact that different realisations of epidemics lead to different spatial configurations for $\{T_{\text{d-r}}\}$. 

Bearing in mind that the population of hosts is homogeneous (i.e. $\alpha$ and $\beta$ do not depend on the d-r pair location) the 
values of transmissibilities for any d-r pair given by Eq.~\eqref{eq_SI:T_maintext} are independent random variables 
 characterized by a
p.d.f., $\rho(T_{\text{d-r}})$. 
Once $\rho(T_{\text{d-r}})$ is available, e.g. numerically, 
 the mean transmissibility can equivalently
 be calculated  as
\begin{equation}
\label{eq_SI:avg_T}
\langle T \rangle = \int_0^1 T \rho(T)\, \text{d}T~.
%\tag{S.6}
\end{equation}
The analysis of $\rho(T_{\text{d-r}})$ is important for understanding the
dependence of $P_{\text{inv}}$ on the degree of synergy and $\langle T \rangle$, as we show below for r- and d-synergy.

\subsection{r-synergy}
\label{sub-sec:r-synergy}

\begin{figure}
\begin{center}

{\includegraphics[clip=true,width=8cm]{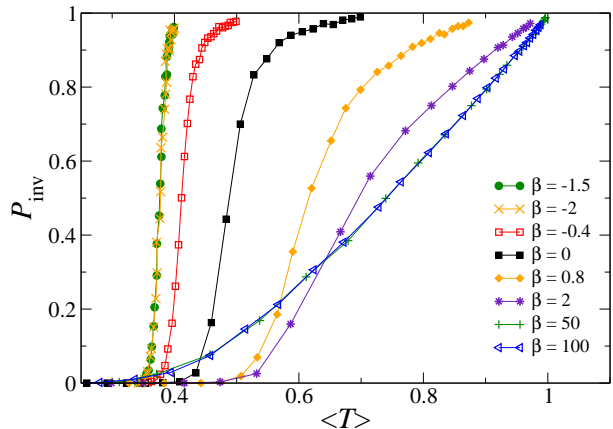}}
\end{center}
\caption{\label{fig:PinvTavg_r-syn}  Probability of invasion as a function
  of the mean  transmissibility for
  r-synergy  in systems of linear size $L=31$. Different curves correspond to different values of
    $\beta$ as marked in the figure.
}
\end{figure}

Fig.~\ref{fig:PinvTavg_r-syn} shows the dependence of $P_{\text{inv}}$
on $\langle T \rangle$ for several values of $\beta$.
For any fixed
value of $\langle T \rangle$, the probability of invasion with
interfering synergy (see the curves corresponding to $\beta<0$ and
marked by open and solid circles
and open squares in Fig.~\ref{fig:PinvTavg_r-syn}) is systematically
greater than for
non-synergistic epidemics (see the curve for $\beta=0$  marked by
solid circles in Fig.~\ref{fig:PinvTavg_r-syn}), i.e. the populations
exhibiting
interfering r-synergy are more vulnerable to invasion than those
without synergy.
For constructive r-synergy, it is possible to distinguish between two
different regimes.
The first regime corresponds to epidemics with moderate synergy (curves for
$\beta=0.8 \; \text{and} \; 2$ in Fig.~\ref{fig:PinvTavg_r-syn}) that are
less invasive than non-synergistic epidemics for all values of
$\langle T \rangle$,
i.e. the curves marked by the solid diamonds and stars are always
below the curve marked by the solid squares.
The second regime corresponds to larger values of $\beta$ (e.g. $\beta=50
\; \text{and} \; 100$ in Fig.~\ref{fig:PinvTavg_r-syn}).
In this case,
there is a range of $\langle T \rangle $ where the curves marked by
open triangles and crosses are above the curve marked by the solid
squares.
In this interval for  $\langle T \rangle$, the epidemics with
constructive r-synergy are more invasive than those  without synergy.

\begin{figure*}
\begin{center}

{\includegraphics[clip=true,width=14cm]{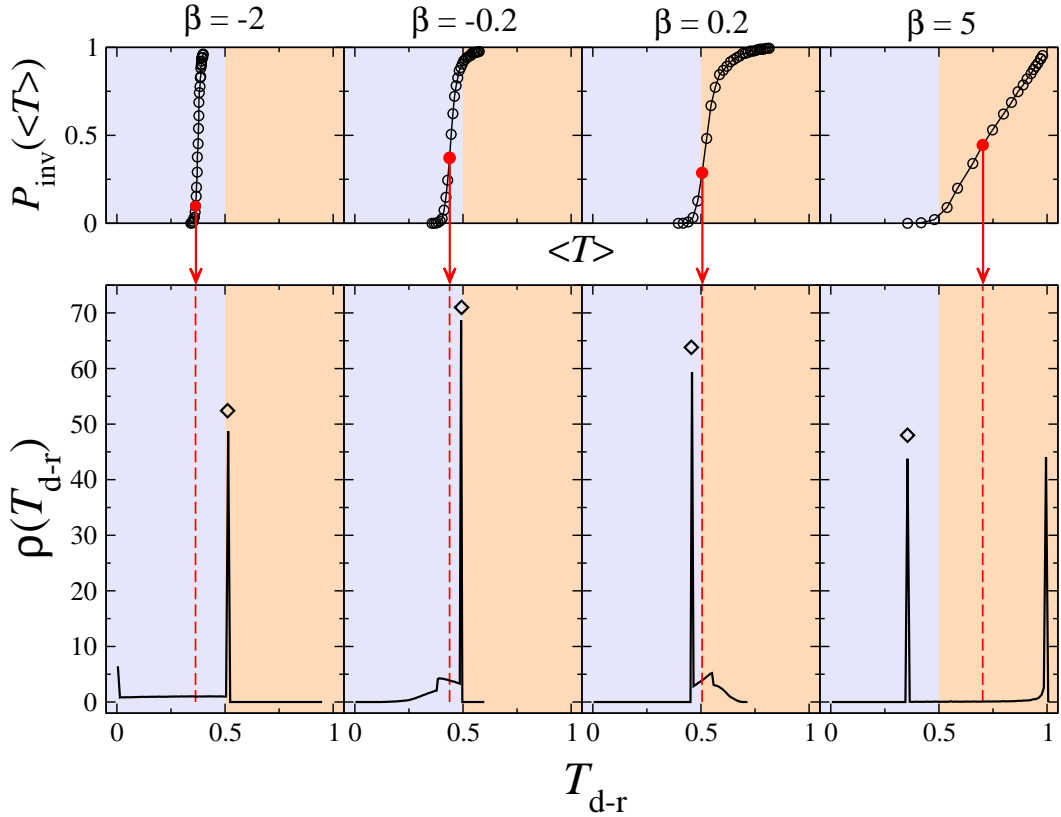}}
\end{center}
\caption{\label{fig:Histo_T_r-syn} The upper panels show
  $P_{\text{inv}}$ vs. $\langle T \rangle$ for four values of $\beta$
  for r-synergy.  The lower panels show the
  p.d.f. $\rho(T_{\text{d-r}})$ corresponding to the points in the
  invasion curves indicated in the upper panels with an arrow. The
  mean of the p.d.f. $\rho(T_{\text{d-r}})$ gives the value of
  $\langle T \rangle$ in the upper panel (Eq.~\eqref{eq_SI:avg_T}).
  The non-synergistic transmissibility $T_1=1-e^{-\alpha \tau}$
  corresponding to the recipients with $n_{\text{r}}=1$ is marked by
  $\diamond$ in all the panels. The critical transmissibility for
  non-synergistic epidemics is $T_c=1/2$. The intervals of transmissibility with
  $T<T_c$ and  $T>T_c$ are indicated by the blue and
  orange shaded regions, respectively.  }
\end{figure*}

These features can be qualitatively illustrated by analysing the
evolution of the distribution of transmissibilities, $\rho(T_{\text{d-r}})$, with
the strength of synergy.
Without synergy, $\rho(T_{\text{d-r}})$ has a
$\delta$-functional shape, $\rho(T_{\text{d-r}})=\delta(T_{\text{d-r}}-T_1)$, where
$T_1=1-e^{-\alpha \tau}$ is the
non-synergistic transmissibility.
The epidemic
is invasive (non-invasive) if $T_1 > T_c=1/2 $ ($T_1 \le
T_c$)~\cite{Grassberger1983}.
Once r-synergy is introduced, the shape of $\rho(T_{\text{d-r}})$
changes.
The
$\delta$-functional peak is still present and it describes recipients
with a single infected neighbour, $n_{\text{r}}=1$.
In addition,
new contributions to $\rho(T_{\text{d-r}})$ appear for $T_{\text{d-r}}
< T_1$ ($T_{\text{d-r}} > T_1$) in case
of negative (positive) values of $\beta$ (see the lower panel in
Fig.~\ref{fig:Histo_T_r-syn} where the $\delta$-functional peak
at $T_{\text{d-r}}=T_1$ is
marked by $\diamond$).
Such contributions come from
recipients surrounded by more than one infected neighbour,
i.e. $n_{\text{r}}>1$.
The number of infected neighbours varies in discrete manner and
this brings a non-smooth functional dependence to
$\rho(T_{\text{d-r}})$ consisting
of cusps associated with the discrete
changes in $n_{\text{r}}(t)$ and smooth components
between cusps 
originated from the continuity of time. 
In other words,
$n_{\text{r}}(t)$ is a piece-wise function of time the integral
  of which produces a
continuous set of transmissibilities
according to Eq.~\eqref{eq_SI:T_maintext}.

For interfering synergy, the average transmissibility is $\langle T
\rangle < T_1$, due to the contribution to $\rho(T_{\text{d-r}})$
of  $T_{\text{d-r}}<T_1$ described above.
If the heterogeneous transmissions associated with
synergy were uncorrelated, the probability of invasion for synergistic
epidemics with the elementary rate $\alpha$ would be smaller than that
corresponding to epidemics with the same value of $\alpha$ but without
synergy.
This is a consequence of the fact that the probability of
invasion for systems with heterogeneous but uncorrelated
transmissions
depends on $\langle T \rangle$ only
\cite{Cox1988,Sander2002,PerezReche_JRSInterface2010}.
However, this is not the case for
epidemics with
interfering synergy which are more invasive than epidemics
  without synergy.
This is due to correlations between synergistic
  transmissibilities.
 Indeed, the d-r pairs that passed the
infection in an epidemic (mostly those with non-synergistic
transmissibility $T_1$) are arranged in a spatially correlated
finger-like manner (as shown in Fig.~3(b) in the
main text) that makes invasion possible.
Therefore, the synergistic epidemics can be mapped onto
  the correlated dynamical bond-percolation problem in which the bond
  probabilities are associated with transmissibilities.

For constructive synergy with moderate values of $\beta$
(corresponding to the first regime mentioned above), most of the
d-r pairs in the system have transmissibility $T_1<T_{\text{c}}=1/2$
(cf. the position of the peak marked by
$\diamond$
with the rest of the distribution  in the panel for
$\beta=0.2$ in Fig.~\ref{fig:Histo_T_r-syn}).
Most of the synergistic
d-r pairs have $T_{\text{d-r}}>T_c$
as can be seen from the comparison of the areas under the curve for
p.d.f. for $T_{\text{d-r}}<T_c $ (excluding non-synergistic
transmissibilities under the peak marked by $\diamond$)
and $T_{\text{d-r}}>T_c $.
However, the abundance and value of the transmissibilities for such
synergistic pairs does not seem to be high enough as to allow for
invasion unless $\langle T \rangle$ is clearly larger than $T_c$.

For larger values of $\beta$, invasion is possible for $\langle T
\rangle < T_{\text{c}}$ because, as shown in
Fig.~\ref{fig:Histo_T_r-syn} for $\beta=5$, the relative number of
synergistic pairs is large enough and they have $T \simeq 1$ so
that the transmission 
of infection is very likely.
Moreover, these pairs are
placed in a spatially correlated manner which also favours the
invasion for $\langle T \rangle < T_{\text{c}}$.

\subsection{d-synergy}
\label{sub-sec:d-synergy}

\begin{figure*}
\begin{center}

{\includegraphics[clip=true,width=14cm]{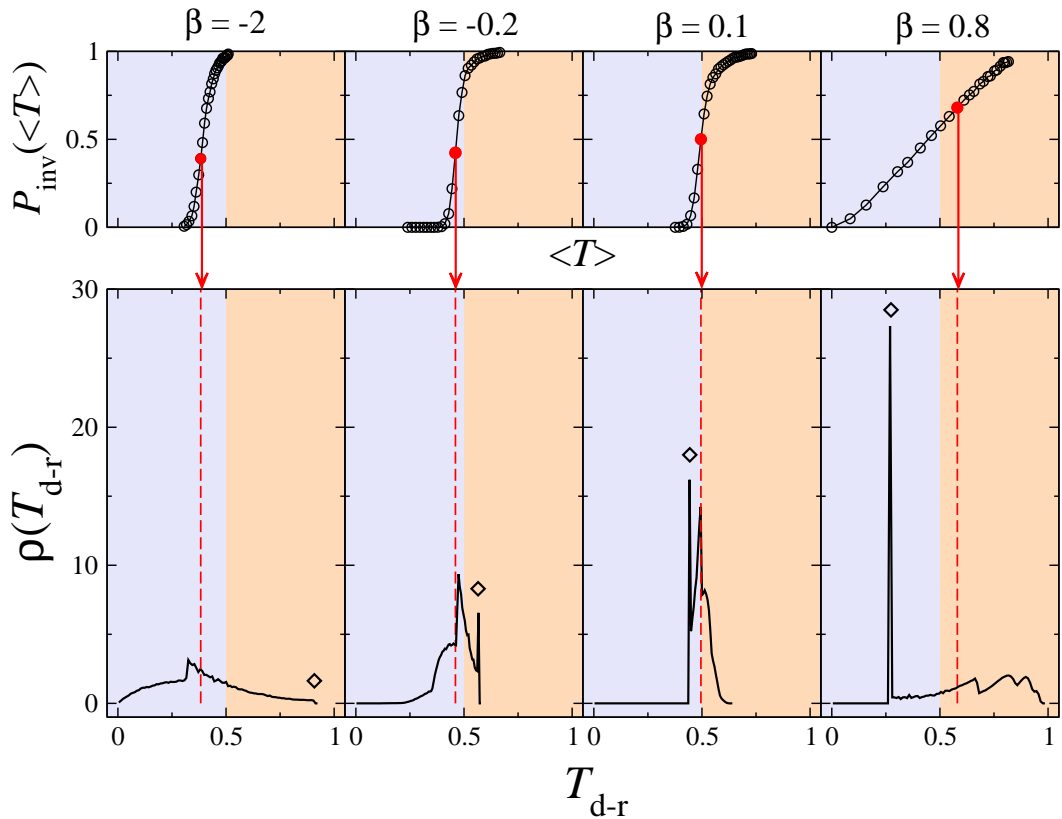}}
\end{center}
\caption{\label{fig:Histo_T_d-syn}
The upper panels show $P_{\text{inv}}$ \emph{vs.}  $\langle T \rangle$ for
four values of $\beta$ for d-synergy.
The lower  panels show the
p.d.f. $\rho(T_{\text{d-r}})$ corresponding to the points in the
invasion  curves indicated in the upper panels with an arrow.
The mean
of the p.d.f. $\rho(T_{\text{d-r}})$ gives the value of $\langle T
\rangle$ in the upper panel (Eq.~\eqref{eq_SI:avg_T}).
The non-synergistic transmissibility $T_0=1-e^{-\alpha
    \tau}$ corresponding to recipients with $n_{\text{d}}=0$
 is marked by $\diamond$ in all the panels. The critical transmissibility for
  non-synergistic epidemics is $T_c=1/2$. The intervals of transmissibility with
  $T<T_{\text{c}}$ and  $T>T_{\text{c}}$ are indicated by the blue and
  orange shaded regions, respectively. 
}
\end{figure*}

The p.d.f.  $\rho(T_{\text{d-r}})$ for epidemics with
d-synergy is shown in Fig.~\ref{fig:Histo_T_d-syn} for several values
of $\beta$.
At  first sight,
the effects of d-synergy on $\rho(T_{\text{d-r}})$ are
qualitatively similar to those of r-synergy.
However, for d-synergy,
the $\delta$-functional peak for the non-synergistic transmissibility $T_{0}$
marked by $\diamond$ corresponds to
 realisations of the epidemic
when the pathogen has not been transmitted from the initially infected host
to any of its neighbours (i.e. the epidemic has not started 
spreading).
Only in this case $n_{\text{d}}=0$ over the whole
infectious period of the initially inoculated host and
$T_{\text{d-r}}=T_0$.
In contrast, $T_{\text{d-r}} \neq T_0$ for every
d-r pair if the epidemic starts, meaning that there is no contribution to
the $\delta$-functional peak from these epidemics.

\begin{figure}
\begin{center}

{\includegraphics[clip=true,width=8cm]{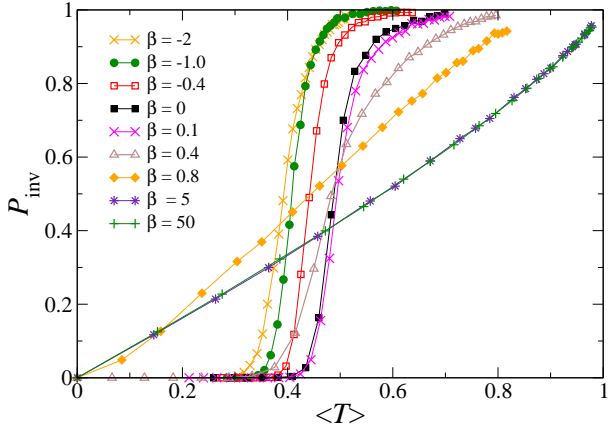}}
\end{center}
\caption{\label{fig:PinvTavg_d-syn}  Probability of invasion as a function
  of the mean transmissibility for
  d-synergy in systems of linear size $L=31$.
Different curves correspond to different values of
    $\beta$ as marked in the figure.
}
\end{figure}

The scenario for epidemics with interfering d-synergy is similar to
that for r-synergy, meaning that invasion can occur for values of
$\langle T \rangle <T_c $
(cf. the curves for $\beta <0$ and
$\beta=0$ in Fig.~\ref{fig:PinvTavg_d-syn}).
For any fixed negative value of
$\beta$, the shift (to the left)
of the invasion curve from the synergy-free one,
$P_{\text{inv}}(\langle T \rangle)$ with
$\beta=0$, is greater 
for d-synergy than for r-synergy.
Correlations in transmission seem to play a very prominent
role for d-synergy since invasion can occur even for very low values
of $\langle T \rangle$.
The fact that d-synergy induces larger shifts for the
$P_{\text{inv}}(\langle T \rangle)$ 
curve towards smaller values of
$\langle T \rangle$ than r-synergy is due to the greater abundance of
synergistic connections for d-synergy, as argued in the main text.
This is clear
from the comparison of the p.d.f. $\rho(T_{\text{d-r}})$ for
$\beta<0$ corresponding to the
two types of synergy shown in Figs.~\ref{fig:Histo_T_r-syn} and
\ref{fig:Histo_T_d-syn}.
The relative number of  
non-synergistic d-r pairs with non-synergistic transmissibility $T_0$ is
always smaller for d-synergy.

For constructive d-synergy, there are two regimes that are
qualitatively similar to those discussed for constructive r-synergy
above.
The regime where invasion is only possible for $\langle T
\rangle > T_c$ exists for very weak constructive synergy 
($0 <  \beta \lesssim
0.1$).
Figs.~\ref{fig:PinvTavg_d-syn} and \ref{fig:Histo_T_d-syn}
illustrate the behaviour in this regime for $\beta=0.1$.
For greater
values of $\beta$, invasion is possible for $\langle T \rangle<T_c = 1/2$
(see the curves for $\beta \geq 0.4$ in Fig.~\ref{fig:PinvTavg_d-syn}).
This is due to the presence of a large number of synergistic
transmissions with $T_{\text{d-r}}>T_c$, 
as illustrated in
Fig.~\ref{fig:Histo_T_d-syn} for $\beta=0.8$.
As argued above, the
peak at $T_0$ is due to epidemics that do not start spreading.
The
rest of contributions to $\rho(T_{\text{d-r}})$
corresponds  to those cases in
which the pathogen is transmitted from the initially infected host to
at least one of its neighbours.
The values of the transmissibilities
for such epidemics are larger and thus $P_{\text{inv}}$ is also
larger. 
In this case, invasion is possible for any positive value of $\langle
T \rangle$ because the epidemic is invasive with high probability
provided it starts spreading.
In other words, many epidemics do not
start spreading for very low value of $\alpha$ (and thus low
transmissibility).
However, the probability that the epidemic starts
spreading is larger than zero for any $\alpha>0$.
Once it starts, invasion almost certainly occurs.
Therefore,
 invasion is possible for any
positive value of $\alpha$, no matter how small the value of $\alpha$ is.

\section{Phase diagram for a simple model} 
\label{sec:phase_diagram}
 
The aim of this section is to present a simple model for evaluation of the phase diagram in $\alpha - \beta$ parameter space for both types of synergy in 
SIR process.

The phase transition from non-invasive to invasive SIR regime in heterogeneous systems with uncorrelated transmissibilities occurs when  
\begin{equation} 
\label{eq:condition}
\langle T \rangle = T_c~ ,   
%\tag{S.7}
\end{equation} 
where $T_c$ is a critical topology-dependent bond-percolation probability  
($T_c=1/2$ for a square lattice) and $\langle T \rangle $ is  
the mean transmissibility~\cite{Sander2002,PerezReche_JRSInterface2010}.  
For a synergistic SIR process, the transmissibilities are heterogeneous due to  variable neighbourhood during infection (bond creating) process and correlated as it follows from our analysis of probability of invasion \emph{vs} mean transmissibilities (see Sec.~\ref{Sec:Corr_Dyn_Perc}).  
Such correlations make exact analytical treatment for synergistic SIR process 
  hardly possible. 
However, if the correlations between transmissibilities are ignored then a  simplified model for synergistic SIR process can be introduced and solved analytically for the boundaries in the phase diagram. 
The simplified assumptions of the model are the following: 
\begin{itemize}
\item[(i)] there are no correlations in transmissibilities;  
\item[(ii)] the neighbourhood of a r-d pair does not change over 
            the infectious period of the donor; 
\item[(iii)] the probabilities of various neighbourhoods of a d-r do not depend on the rates $\alpha$ and $\beta$;
\end{itemize} 

Under these assumptions, the analytical expressions for the phase boundaries reproducing qualitatively all the features found numerically can be derived and analysed. 
 
\subsection{d-synergy} 

We start analysis of the simplified model for synergistic SIR process on a square lattice (for concreteness) with the case of d-synergy.  
Let us consider an infected node (other than the initially infected host) 
which attempts to transmit infection (create a bond) to one of its susceptible neighbours during its infectious period $\tau = 1$.  
This process can occur with different probability depending on the number of  infected nodes linked to the infecting one (donor).  
Applying assumptions (i)-(ii) that transmissibilities are uncorrelated and a certain random  neighbourhood does not change over the infectious period of the donor, we can calculate the mean transmissibility  in the following way, 
\begin{equation} 
\label{eq:mean_T_d} 
\langle T \rangle = \sum_{n=1}^3 p_n T_n~.  
%\tag{S.8}
\end{equation} 
Here $p_n$ is the probability that the donor is connected to 
$n$ ($n=1,2,3$ for square lattice) infected nodes and $n$-dependent transmissibility, $T_n$, for the d-r pair is 
\begin{equation} 
\label{eq:Tn_d}
 T_n =  \left\{
\begin{array}{ccc}
 1-e^{-\alpha -n\beta} ~~~&\text{if}&~~~\alpha +n\beta>0
 \\ 
 0 ~~~&\text{if}&~~~\alpha +n\beta\le0~. 
\end{array}
\right.
%\tag{S.9}
\end{equation} 
Different neighbourhoods of the d-r pair occur with
 probabilities $p_n$.  
Under assumption (iii), these probabilities do not depend on $\alpha$ and $\beta$ and given by the following expressions:  
\begin{align} 
 p_1 &= \frac{3 p (1-p)^2}{1-(1-p)^3}~, 
\nonumber \\ 
p_2 &=  \frac{3 p^2 (1-p)}{1-(1-p)^3}~,
\nonumber \\ 
p_3 &=  \frac{p^3}{1-(1-p)^3}~, 
\label{eq:pn_d} 
%\tag{S.10}
\end{align}
where $p_n$ is the probability that the infecting node is linked to $n$ nodes  given that it is linked at least to one out of three possible nodes and  
 $p$ is the bond probability with $p=T_c$ at criticality.  
 In Eq.~\eqref{eq:pn_d}, we used the assumption (i) that the bonds were created independently.  
 
The locus of critical points can be found by solving Eq.~\eqref{eq:condition}, 
\begin{equation} 
 T_c = \sum_{n=1}^3p_n(T_c)T_n~,  
\label{eq:Tc_d_2} 
%\tag{S.11}
\end{equation} 
where $\langle T \rangle$ is given by Eqs.~\eqref{eq:mean_T_d}-\eqref{eq:pn_d}. 
Eq.~\eqref{eq:Tc_d_2} can be recast as $(\alpha > 0)$,  
\begin{widetext}
\begin{equation} 
\alpha = \begin{cases} 
 -\ln(1-T_c) + \ln\left(p_1(T_c)e^{-\beta}+p_2(T_c)e^{-2\beta} + 
p_3(T_c)e^{-3\beta}\right)~,~~~&\text{if}~~~\beta > -\alpha/3~,  
%\label{eq:solution_d_1} 
%\tag{S.12}
\\ 
 -\ln(1-T_c-p_3(T_c)) + \ln\left(p_1(T_c)e^{-\beta}+p_2(T_c)e^{-2\beta}\right)~,~~~&\text{if}~~~-\alpha/2 < \beta \le  -\alpha/3~,  
%\label{eq:solution_d_2} 
%\tag{S.13}
\\  
 -\ln(1-T_c-p_3(T_c)-p_2(T_c)) + \ln\left(p_1(T_c) \right) -\beta ~,~~~&\text{if}~~~-\alpha < \beta \le  -\alpha/2~,  
%\label{eq:solution_d_3}
%\tag{S.14}
\end{cases}
\label{eq:solution_d_1} 
%\tag{S.12}
\end{equation}
\end{widetext}

 and
\begin{equation}  
\langle T \rangle = 0 ~,~~~\text{if}~~~ \beta \le  -\alpha~. 
\label{eq:solution_d_4} 
%\tag{S.13}
\end{equation} 
i.e. the system is in non-invasive regime. 
  
In the limiting case of small synergy, $|\beta| \ll 1$, the locus of critical points is given by a stright line,  
\begin{equation} 
 \alpha \simeq -\ln(1-T_c) - \langle n_b \rangle \beta = -\ln(1-T_c) - \sum_{n=1}^3 np_n\beta~,  
\label{eq:small_beta_d} 
%\tag{S.14}
\end{equation}
where $\langle n_b \rangle $ is the mean number of bonds attached to the donor given at least one bond atttached.  
For square lattice with $T_c=1/2$ and  
$p_1=p_2=3/7$, $p_3=1/7$, Eq.~\eqref{eq:small_beta_d} gives  
\begin{equation} 
 \alpha \simeq \ln(2) - 12\beta/7~.   
\label{eq:small_beta_d_1} 
%\tag{S.15}
\end{equation} 

In the limiting case of strong interference, $\beta \to -\infty$,
  the phase boundary approaches the straight line asymptote, given by 
\begin{equation}
\begin{split} 
 \alpha &= -\ln(1-T_c-p_3(T_c)) + \ln\left(p_1(T_c)e^{-\beta}+p_2(T_c)e^{-2\beta}\right) \\
%\nonumber \\  
&\simeq \ln \frac{p_2(T_c)}{1-T_c-p_3(T_c)}-2\beta = \ln(6/5) -2\beta ~,   
\label{eq:small_beta_d_2} 
%\tag{S.16}
\end{split} 
\end{equation}
for square lattice. 

\begin{figure}%[h] 
\begin{center} 
{\includegraphics[clip=true,width=8cm]{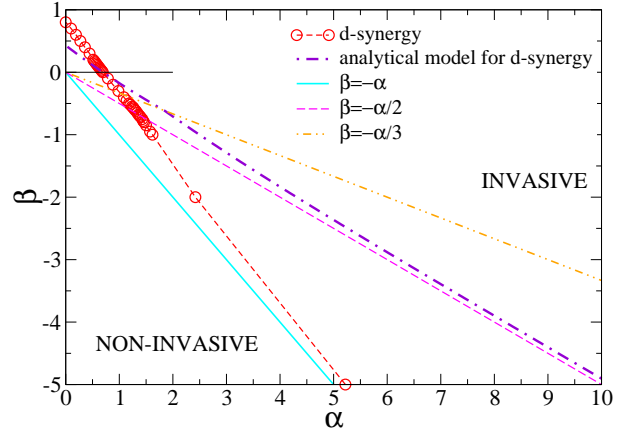}} 
\end{center} 
\caption{\label{fig:d_synergy2}Locus of critical points for d-synergy within simple analytical model (dot-dashed line) compared with the exact numerical data (dashed line marked by circles). 
The intersections of the  critical lines  with straight lines,  $\beta=-\alpha/n$, 
indicate the location of the kinks on the phase boundaries.  
} 
\end{figure}

The locus of critical points for d-synergy given by 
 Eqs~\eqref{eq:solution_d_1} and \eqref{eq:solution_d_4} 
 is shown in Fig.~\ref{fig:d_synergy2} (dot-dashed line).  
The intersections of the critical line with the straight lines, 
$\beta= -\alpha/n $ ($n=1,2,3$), correspond to the changes in the regimes for the infection rates in Eqs.~\eqref{eq:solution_d_1} and  \eqref{eq:solution_d_4} and thus lead to appearance of the kinks (discontinuities in the derivatives) on the critical line.  
Within our approximations, the values of $p_n$ for square lattice are such that there is only one kink on the phase boundary for simple analytical model 
(corresponding to the crossing point of the phase boundary with $\beta= -\alpha/3 $) 
and in the asymptotic regime, $\beta \to -\infty$, the phase boundary  approaches linear asymptote, $\beta \simeq \beta'_0 -\alpha/2$ 
 (with positive constant $\beta'_0>0$).  

Kinks are also expected on the phase boundary for the exact model. In order to check this, we have analysed the behaviour of the exact phase boundary obtained numerically around the point of intersection with the line $\beta=-\alpha/2$ (cf. Fig.~\ref{fig:d_synergy2}). We have tested the presence of a kink by fitting a linear dependence $\beta = a+b \alpha$ to the data above and below the intersection. This procedure reveals a significant difference in the slope which takes values $b=-0.97$ and $b=-1.26$ above and below the intersection, respectively.
 
%\begin{figure}[h] 
%\begin{center} 
%{\includegraphics[clip=true,width=12cm]{Toy_model/d_synergy1.eps}} 
%\end{center} 
%\caption{\label{fig:d_synergy1}Locus of critical points for d-synergy within %simple analytical model. The intersections of the curves  $\beta=-\alpha/n$ %with the critical line (solid) indicate the location of the kinks on the %critical line.  
%} 
%\end{figure} 

The locus of critical points in $\alpha-\beta$ parameter space 
derived within the simple analytical model (dot-dashed line in Fig.~\ref{fig:d_synergy2}) is similar in shape to that obtained 
 numerically (dashed line marked by the circles in Fig.~\ref{fig:d_synergy2}). 
However, due to simplifying approximations (i)-(iii), 
the analytical model does not capture the asymptotic behaviour, $\beta \simeq \beta_0 -\alpha$   (with positive constant $\beta_0>0$), obtained numerically for $\beta\to\-\infty$. 
This could be due to the fact that the actual probabilities $p_n$ 
(possibly depending on $\alpha$ and $\beta$ might be quite different from those used in the analytical model (see Eqs.~\eqref{eq:pn_d}).  
The difference in gradients in the small-$\beta$ limit between 
analytically and numerically found phase boundaries can be due to similar reasons. 
 
\subsection{r-synergy} 
 
For $r$-synergy, the transmission of infection from a donor to a recipient can occur in the presence of different number of infected neighbours 
(in addition to the donor) of the recipient 
(for concreteness, we consider a  square lattice for which $n=1,2,3$).  
Under assumptions (i) and (ii), the mean transmissibility for the d-r pair  is, 
\begin{equation} 
\langle T \rangle = \sum_{n=0}^3 q_n T_n~,   
\label{eq:r_synergy} 
%\tag{S.17}
\end{equation} 
where $q_n$ is the probability that the recipient has $n$ infected neighbours (in addition to the donor) and synergistic transmissibilities are given by  
Eq.~\eqref{eq:Tn_d} where $n=0,1,2,3$.  
The probabilities $q_n$ of different neighbourhoods of the recipient can be defined through the probability, $t$ (parameter of the model), that 
 a nearest neighbour of  the recipient host (different from the donor) 
is in the infected state,  
\begin{equation}  
q_n = C_n^3 t^n(1-t)^{3-n} 
~, 
\label{eq:qn_r} 
%\tag{S.18}
\end{equation} 
where we used the assumption about the independence of infection events for different  neighbours of recipient.   
Given assumption (iii), from Eq.~\eqref{eq:condition} written in the form,  
\begin{equation}
T_c = \sum_{n=0}^3 q_n T_n 
~, 
\label{eq:critical_r} 
%\tag{S.19}
\end{equation}
we can easily obtain the following equation for the locus of critical points:  
\begin{widetext}
\begin{equation}
\alpha =
\begin{cases}
-\ln(1-T_c) + \ln\left(q_0(t) + q_1(t)e^{-\beta}+q_2(t)e^{-2\beta} + 
q_3(t)e^{-3\beta}\right) 
~,& ~~~\text{if}~~~\beta > -\alpha/3  
\\ 
-\ln(1-T_c-q_3(t)) + \ln\left(q_0(t) + q_1(t)e^{-\beta}+q_2(t)e^{-2\beta}  
\right) 
~, & ~~~\text{if}~~~-\alpha/2 < \beta \le  -\alpha/3  
\\  
-\ln(1-T_c-q_3(t)-q_2(t)) + \ln\left(q_0(t) + q_1(t)e^{-\beta} \right) 
~, &~~~\text{if}~~~-\alpha < \beta \le  -\alpha/2  
\\  
\alpha_{*}=\ln\frac{q_0(t)}{q_0(t)-T_c}   ~, & ~~~\text{if}~~~ \beta \le  -\alpha~.
\end{cases}
\label{eq:solution_r_1} 
%\tag{S.20}
\end{equation}
\end{widetext}
%%\begin{subequations} 
%%\label{eq:solution_r_1} 
%% \tag{S.22}
%\begin{align}
%\alpha &= -\ln(1-T_c) + \ln\left(q_0(t) + q_1(t)e^{-\beta}+q_2(t)e^{-2\beta} + 
%q_3(t)e^{-3\beta}\right) 
%~,~~~\text{if}~~~\beta > -\alpha/3  
%\label{eq:solution_r_1} 
%\tag{S.20}
%\\ 
%&= -\ln(1-T_c-q_3(t)) + \ln\left(q_0(t) + q_1(t)e^{-\beta}+q_2(t)e^{-2\beta}  
%\right) 
%~,~~~\text{if}~~~-\alpha/2 < \beta \le  -\alpha/3  
%\label{eq:solution_r_2} 
%\tag{S.23}
%\\  
%&= -\ln(1-T_c-q_3(t)-q_2(t)) + \ln\left(q_0(t) + q_1(t)e^{-\beta} \right) 
%~,~~~\text{if}~~~-\alpha < \beta \le  -\alpha/2  
%\label{eq:solution_r_3} 
%\tag{S.24}
%\\  
%&=\alpha_{**}=\ln\frac{q_0(t)}{q_0(t)-T_c}   ~,~~~\text{if}~~~ \beta \le  -\alpha~.
%\label{eq:solution_r_4} 
%\tag{S.25}
%\end{align}
%%\end{subequations}
 
In the limit of large values of $\beta \to \infty$, the value of $\alpha$ approaches  the asymptotic value  $\alpha_c(\infty)$,
\begin{equation}  
\alpha \to  \alpha_c(\infty) =\ln \frac{q_0}{1-T_c} 
~. 
\label{eq:large_beta_r} 
%\tag{S.21}
\end{equation} 
The value of  $\alpha_c(\infty)$ can be both positive and negative.  
We know from exact numerical analysis that $\alpha_c(\infty)>0$ ($\alpha_c(\infty)\simeq 0.2$) and thus we assume that  
\begin{equation}  
q_0 = (1-t)^3 > 1-T_c 
~, 
\label{eq:q0_r} 
%\tag{S.22}
\end{equation} 
i.e $t<1-(1-T_c)^{1/3}$ ($t<7/8$, for square lattice).  
 
In the limiting case of small values of $|\beta| \to 0$,  
\begin{equation}  
\begin{split}
\alpha \to \alpha_c(\infty)= &- \ln(1-T_c) - \sum_1^3 nq_n \beta = \\ 
= &- \ln(1-T_c) -  \langle n \rangle \beta 
~,
\end{split}
\label{eq:small_beta_r} 
%\tag{S.23}
\end{equation} 
where $\langle n \rangle$ is the mean value of infected neighbours 
(excluding the donor) of  the  recipient.  
  
\begin{figure}[b] 
\begin{center} 
{\includegraphics[clip=true,width=8cm]{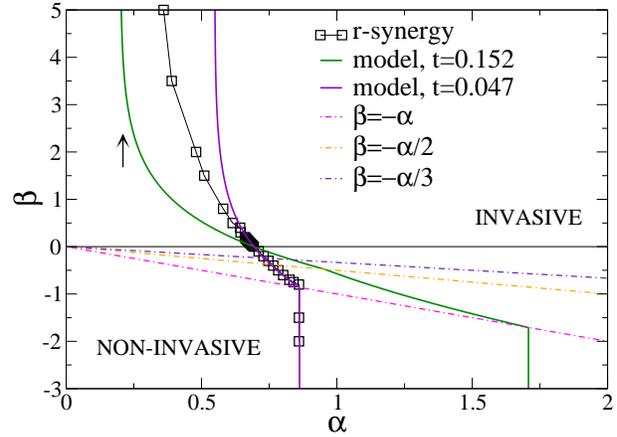}} 
\end{center} 
\caption{\label{fig:r_synergy2}Locus of critical points for r-synergy within
  simple analytical model for two values of parameter $t$ 
(as indicated in the legend) 
compared with the exact numerical data. 
The intersections  of the critical lines 
with the straight lines  $\beta=-\alpha/n$ 
 indicate the locations of the kinks on the phase boundaries.  
} 
\end{figure}

In the limiting case, $\beta \to -\infty$, the behaviour of the analytical critical line depends on the value of $t$. 
In particular, if 
$q_0=(1-t)^3 > T_c$ then the critical line intersects the straight lines 
$\beta = -\alpha/n$ (resulting in kinks on the critical line)  for all values of $n=1,2,3$ and becomes a vertical border at $\alpha=\alpha_{*}$ for $\beta \le -\alpha_{*}$. 
Numerical data support such a scenario with the vertical border. 
The positions of both asymptotic value of $\alpha_c(\infty)$ and  vertical border 
$\alpha_{*}$ vary with the value of $t$. 
In Fig.~\ref{fig:r_synergy2}, the analytical critical lines are shown for two
values of the node occupation probability $t$. 
For $t=0.047$, the behavior around small values of $\beta$ and position of the
vertical border found numerically are reproduced quite well by the analytical
model. 
However, the analytical curve strongly deviates from the numerical one for
$\beta \gg 1$ and fails to reproduce the value of $\alpha_c(\infty)$. 
If we try to mimic the value of $\alpha_c(\infty)$ by tuning $t$ (see the curve for
$t=0.152$), 
then the gradient at small $|\beta|$ and position of the vertical border are
significantly off the numerical values. 
Such deviations area consequence of approximations (i)-(iii) used in the analytical model. 

Overall, comparing the two types of synergy within the simple analytical model we can
conclude that the main differences between them come from the presence of non-synergistic
transmission events that are possible for r-synergy with any value of $\beta$ when the recipient is challenged by a single infected neighbour. 
Such transmission events with transmissibility $\propto 1-e^{-\alpha}$ 
are responsible for the appearance of the asymptotic value
$\alpha_c(\infty)$ 
and vertical border $\alpha_{*}$  for r-synergy. 

%%% BibTex bibliography %%%%%%%%%%
%\bibliographystyle{apsrev}
%apsrev4-1.bst     - BibTeX styles for use for Phys. Rev. journals
%apsrev4-1long.bst - Same as above, but shows titles for cited journal articles
%\bibliographystyle{apsrev4-1.bst}
%\bibliography{jons,Chris,../bibliography/bibliography}

%Merlin.mbs v4.21 2009-07-09.
%

\end{document}